\documentclass[
  11pt,
  captions=centeredbeside,
  letterpaper,
  headings=optiontoheadandtoc,
  numbers=endperiod,
  oneside,
]{scrartcl}

\usepackage[
    autooneside=false,
    automark,
    ]{scrlayer-scrpage}
\usepackage{setspace}
\usepackage{rotating}

\DeclareUnicodeCharacter{0301}{\'{e}}

\ohead{\leftmark}
\setkomafont{pageheadfoot}{\color{darkgray}\small\sffamily\upshape}
\setkomafont{pagenumber}{\color{darkgray}\small\sffamily\upshape}
\setkomafont{captionlabel}{\bfseries\sffamily
}
\setkomafont{caption}{\normalfont\sffamily
}
\setcapindent{0pt}
\RedeclareSectionCommand[
  beforeskip=0.75\baselineskip,
  afterskip=0.25\baselineskip,
  afterindent=false,
]{section}
\RedeclareSectionCommand[
  beforeskip=0.50\baselineskip,
  afterskip=0.25\baselineskip,
  afterindent=false,
]{subsection}
\deffootnote[0.5em]{1em}{1em}{\textsuperscript{\thefootnotemark}}

\usepackage[margin=1in]{geometry}
\usepackage{enumitem}
\setlist{nosep}

\usepackage{stackengine}
\usepackage{anyfontsize}
\usepackage{lineno}
\usepackage{microtype}
\usepackage{amssymb}
\usepackage{amsmath}
\usepackage[normalem]{ulem}
\usepackage{xspace}
\usepackage{verbatim}
\usepackage{multirow}
\usepackage{natbib}
\usepackage{enotez}
\DeclareInstance{enotez-list}{plain}{paragraph}{
    heading={},
    number = \textsuperscript{#1},
    notes-sep=0pt,
}

\usepackage[
    usenames,
    dvipsnames,
]{xcolor}
\usepackage{graphicx}
\usepackage[
  colorlinks=true,
  linkcolor=MidnightBlue,
  citecolor=MidnightBlue,
  urlcolor=MidnightBlue,
  ]{hyperref}
\hypersetup{pdftitle={VLA/VLBA to ngVLA Transition Advisory Group Report}}
\usepackage{tikz}
\usetikzlibrary{calc,fadings}
\tikzstyle{every picture}+=[remember picture,overlay,inner sep=0pt]

\usepackage[
    capitalize,
]{cleveref}
\crefformat{section}{\S#2#1#3}
\crefrangeformat{section}{\S\S#3#1#4--#5#2#6}
\crefname{section}{\S}{\S\S}

\usepackage[
    detect-all=true,
    range-units=single,
]{siunitx}

\DeclareSIUnit{\GHz}{GHz}

\begin{document}

%%%%%%%%%%%% Cover Art
\definecolor{twilight}{rgb}{0.0,0,0.1}
\pagecolor{twilight}
\thispagestyle{empty}

\begin{figure}
\vbox{
\includegraphics[width=\textwidth]{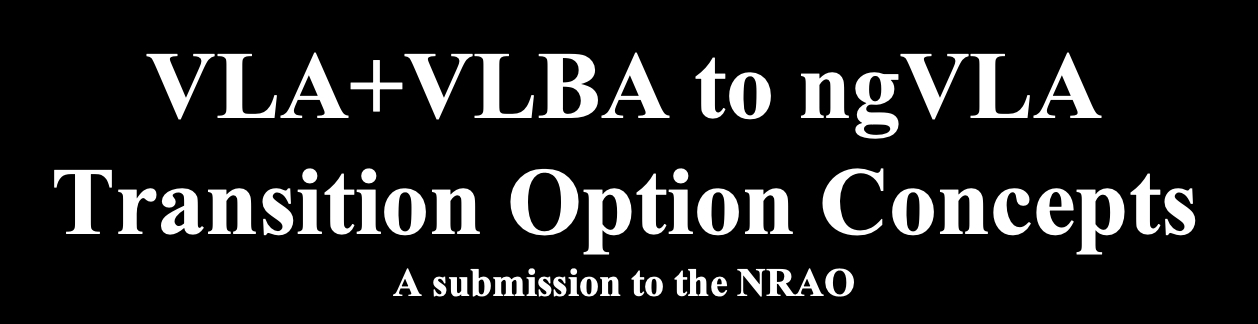}
\includegraphics[width=\textwidth]{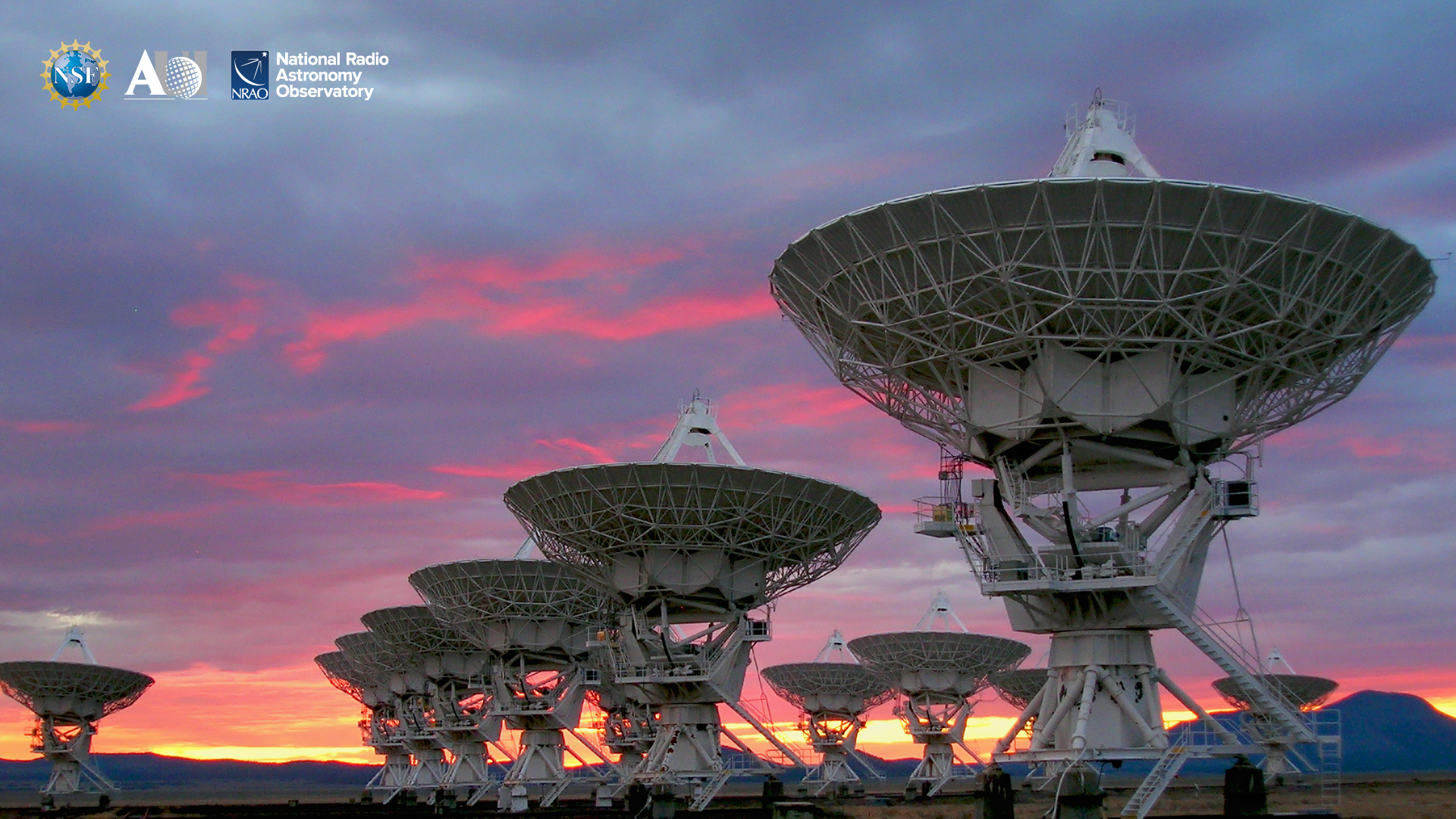}
\includegraphics[width=\textwidth]{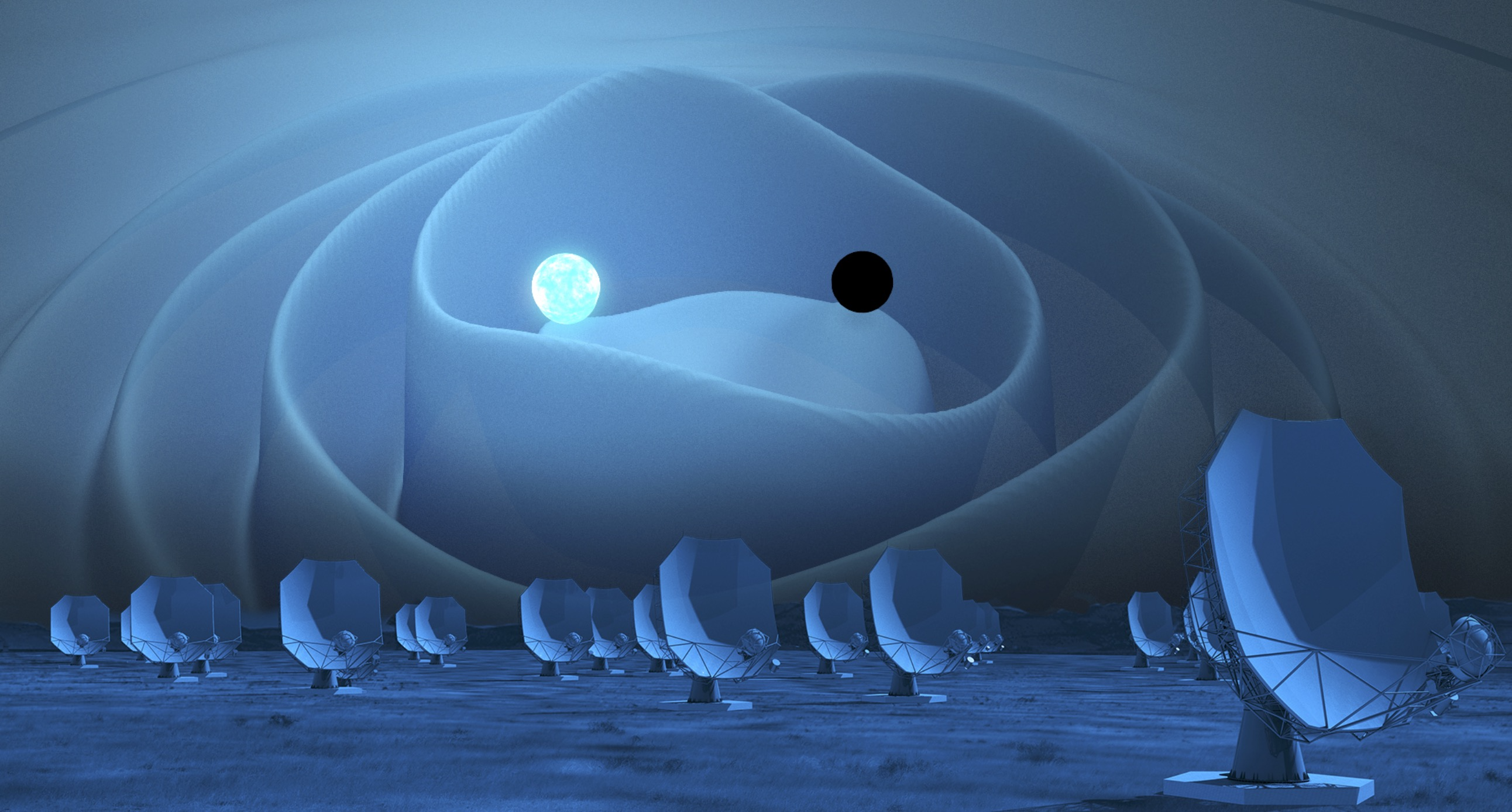}
}
\end{figure}

\clearpage

\pagecolor{white}
\renewcommand{\thempfootnote}{\arabic{mpfootnote}}
\vspace{\baselineskip}
\setfootnoterule[0pt]{10em}
\deffootnote[0em]{1em}{1em}{\textsuperscript{\thefootnotemark}}

\large{
\begin{center}
\textbf{TAG Co-chairs}:
\mbox{Alessandra Corsi,\endnote{William H.~Miller~III Department of Physics and Astronomy, Johns Hopkins University}\label{aff1}}
\mbox{T.~Joseph~W.~Lazio\endnote{Jet Propulsion Laboratory, California Institute of Technology}\label{aff2}}\\

\textbf{TAG Members:}
\mbox{Stefi Baum,\endnote{Dept.\ of Physics \& Astronomy, Univ.\ of Manitoba, Canada}\label{aff3}}
\mbox{Simona Giacintucci,\endnote{U.{}S.\ Naval Research Laboratory}\label{aff4}}
\mbox{George Heald,\endnote{CSIRO Space \& Astronomy, Australia}\label{aff5}}
\mbox{Patricia Henning (\textit{ex officio}),\endnote{National Radio Astronomy Observatory}\label{aff6}}
\mbox{Ian Heywood,\endnote{Dept.\ of Physics, Oxford Univ.; South African Radio Astronomy Observatory}\label{aff7}}
\mbox{Daisuke Iono,\endnote{National Astronomical Observatory of Japan}\label{aff8}}
\mbox{Megan Johnson,\endnote{U.{}S.\ Naval Observatory}\label{aff9}}
\mbox{Michael\,T.\,Lam,\endnote{SETI Institute}\label{aff10}}
\mbox{Adam Leroy,\endnote{Dept.\ of Astronomy,
The Ohio State University}\label{aff11}}
\mbox{Laurent Loinard,\endnote{Institute of Radio Astronomy and Astrophysics,
National Autonomous University of Mexico}\label{aff12}}
\mbox{Leslie Looney,\endnote{Department of Astronomy, University of Illinois}\label{aff13}}
\mbox{Lynn Matthews,\endnote{MIT Haystack Observatory}\label{aff14}}
\mbox{Ned Molter,\endnote{Dept.\ Astronomy,
University of California at Berkeley }\label{aff15}}
\mbox{Eric Murphy (\textit{ex officio}),\endnote{National Radio Astronomy Observatory}\label{aff6}}
\mbox{Eva Schinnerer,\endnote{Max Planck Institute for Astronomy}\label{aff17}}
\mbox{Alex Tetarenko,\endnote{Dept.\ of Physics \& Astronomy, Univ.\ of Lethbridge, Canada}\label{aff18}}
\mbox{Grazia Umana,\endnote{INAF Osservatorio Astrofisico di Catania, Italy}\label{aff19}}
\mbox{Alexander van der Horst\endnote{Dept.\ of Physics, The George Washington University}\label{aff20}}
\end{center}}

\printendnotes*[]

\vspace{2\baselineskip}
\noindent
\textsf{Correspondence:} 
%\href{mailto:ngvla-tag@listmgr.nrao.edu}{\texttt{ngvla-tag@listmgr.nrao.edu}}

\vspace{\baselineskip}
\noindent
\textsf{Cover art:} \emph{Photo of the \hbox{VLA}; Artistic rendering of the \hbox{ngVLA}. (Images' Credits: NRAO/AUI/NSF)} 

\vspace{\baselineskip}
\noindent
\textsf{Acknowledgements}%
\emph{
  A.{}C.\ acknowledges support from the National Science Foundation, the National Aeronautics and Space Administration, and the U.{}S.\ Department of Energy. Part of this research was carried out at the Jet Propulsion Laboratory,
California Institute of Technology, under a contract with the National
Aeronautics and Space Administration. L.L. acknowledges the support of UNAM-DGAPA PAPIIT grants IN112820 and IN108324, and CONACYT-CF grant 263356. 
Basic research in radio astronomy at the Naval Research Laboratory is supported by 6.1~Base funding. 
The NANOGrav Collaboration receives support from National Science
Foundation Physics Frontiers Center award numbers 1430284
and~2020265.
The National Radio Astronomy Observatory is a facility of the National Science Foundation operated
under cooperative agreement by Associated Universities, Inc. 
}

%%%%%%%%%%%% Table of Contents
\clearpage

\section*{Executive Summary}\label{sec:execsumm}

The next-generation Very Large Array (ngVLA) is intended to be the premier centimeter-wavelength facility for astronomy and astrophysics, building on the substantial scientific legacies of the Karl G.~Jansky Very Large Array (VLA) and the Very Long Baseline Array (VLBA).
The ngVLA would open a new window on the Universe through ultra-sensitive imaging of thermal line and continuum emission to milliarcsecond resolution, while delivering unprecedented broad-band continuum imaging and polarimetry of non-thermal emission.
The ngVLA would provide a critical electromagnetic complement to a suite of particle detectors and gravitational-wave observatories, as well as space- and ground-based telescopes operating from infrared to gamma-ray wavelengths, hence enabling multi-messenger and multi-band astronomy and astrophysics.

Current construction plans call for the ngVLA to leverage some of the physical infrastructure of both the VLA and the \hbox{VLBA}, potentially drawing on overlapping personnel and information infrastructure.
Multiple options can be envisioned for a VLA+VLBA to ngVLA transition.  
At one extreme, the VLA and VLBA could be kept operating at close-to-maximum scientific capability as they are now while the ngVLA is constructed, until the ngVLA reaches science capabilities comparable to those of the current \hbox{VLA+VLBA}.
The benefit of this approach is that it maximizes scientific opportunities during the transition; its disadvantage is the likely higher effective construction cost (as the \hbox{VLA}, \hbox{VLBA}, and early ngVLA operations costs must all be supported). 
At another extreme, the transition could be as rapid as possible, essentially shutting down the VLA and VLBA entirely to be replaced by the ngVLA.  This option might be the most cost-effective, but it would imply 
a clear, unacceptable  loss of scientific training, productivity, and opportunity.

In order to assess risks and benefits of possible transition plans in between the above two extremes, the ngVLA project established the VLA+VLBA to ngVLA Transition Advisory Group (TAG)---a group of~18 members of the U.{}S.\ and international astronomical community (including members of the ngVLA Science Advisory Council).
The TAG was charged with identifying
(i)~scientific opportunities in the coming decade that will critically benefit from complementary and/or unique observations at radio wavelengths;
(ii)~key stakeholders for the transition;
(iii)~parameters of interest by which to compare transition options; and
(iv)~necessary metrics by which to quantify the impact of different transition options.
The primary deliverable from the TAG is a ``VLA+VLBA to ngVLA Transition Option Concepts'' report that includes a prioritized list of transition options.

The TAG assembled a set of VLA-VLBA-ngVLA science use cases, leveraging a variety of resources including the \textit{ngVLA Science Book}, NRAO press releases, and, ultimately, the \textit{Pathways to Discovery} Decadal Survey.
The science use cases were augmented by discussion of the VLA in the \textit{Origins, Worlds, and Life} (Planetary Science \& Astrobiology) Decadal Survey and the \textit{Exploring and Safeguarding Humanity's Home in Space} (Solar \& Space Physics) Decadal Survey.
These science cases, and particularly those identified as ones to which the VLA or VLBA or both would make a ``very significant'' contribution (as judged by \textit{Pathways to Discovery}), were used to assess the scientific impact of a number of technical options provided by an NRAO Internal Technical Analysis Team (ITAT). Considerations of transition options were limited to hardware, and did not encompass other critical aspects of the VLA or VLBA such as the implementation of scientific operations, data processing, and data products delivery to the community.
The NRAO also provided a nominal plan for the design, development, and construction of the \hbox{ngVLA}, which envisions a 10~year construction interval with early science beginning three years after the start of construction.

After internal discussion, discussion with the \hbox{NRAO}, and after receiving further feedback by the NRAO \hbox{ITAT} on potential cost and technical/personnel impacts of various transition options, the TAG concluded the following. A reasonable transition plan  starts with a three-year interval during the initial ngVLA construction when VLA capabilities remain consistent with current capabilities; this initial phase is followed by a two-year interval during which one or both of the transition options described below could be used concurrently with ngVLA Early Science:
\begin{itemize}
 \item The VLA receiver suite is reduced at each antenna, provided that at least five of the current frequency bands are maintained at all antennas, with the notional set being L-, S-, C-, X-,and K~bands.
 \end{itemize}
 Should the transition option above be infeasible or insufficient, an additional option is to:
 \begin{itemize}
\item Adopt a fixed VLA configuration, recognizing that no single configuration has been identified that preserves a sufficient range of capabilities in angular resolution, flux density sensitivity, and surface brightness sensitivity.
\end{itemize}
\noindent%
The TAG recommends further that during the transition from the VLA+VLBA to the ngVLA, the VLBA observational capabilities remain unchanged compared to current VLBA capabilities, including the full receiver suite.

\textit{An unavoidable conclusion of the TAG's assessment is that any reduction in capability to the VLA or VLBA will have a reduced science return.}
Hence, the \textit{preliminary} transition plan described here presents a reduction that would enable the VLA and VLBA to conduct high-profile science during the construction of the \hbox{ngVLA} in the context of a reduced science portfolio. \textit{This reduction is acceptable only as a temporary measure to enable the enhanced scientific return that the ngVLA construction would bring.}

\clearpage
\addtokomafont{pagenumber}{\color{lightgray}}

\noindent\textit{VLA/VLBA to ngVLA TAG Report} \\
Technical Report %\href{add link}{add link} 
\\
2025 January

\setcounter{tocdepth}{2}
\tableofcontents
\clearpage

%%%%%%%%%%%% Introduction

\section{Introduction} \label{sec:intro}

The next-generation Very Large Array (ngVLA) project of the National Radio Astronomy Observatory (NRAO) is intended to be the premier centimeter-wavelength facility for astronomy and astrophysics, building on the substantial scientific legacies of the Karl G.~Jansky Very Large Array (VLA) and the Very Long Baseline Array (VLBA).

\begin{figure}[bh]
  \centering
\includegraphics[width=\textwidth]{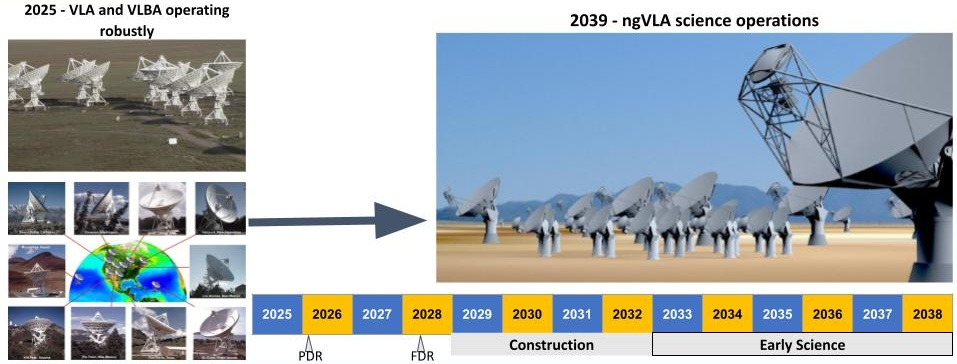}
  \caption{{This report provides scientifically-based recommendations for the transition from the currently-operational VLA and VLBA to the ngVLA, as developed by the TAG.  At the time of writing, the timeline for the ngVLA is for there to be a System Preliminary Design Review (PDR) and a System Final Design Review (FDR), followed by construction beginning in~2029 and Early Science in~2033.  This timeline influenced the discussions of the TAG.  Image credits: NRAO / Associated Universities Inc. (AUI) / National Science Foundation (NSF).}}
  \label{fig:transition}  
\end{figure}

The current construction plans envision that some fraction of the physical infrastructure of both the VLA and the \hbox{VLBA} will be leveraged in order to obtain the \hbox{ngVLA}.
This would have the potential benefit that the construction cost of the ngVLA should be lower than it otherwise would be.
Adopting this approach would require a well-defined process to the transition from the currently-functioning VLA and VLBA to the future \hbox{ngVLA} (Figure~\ref{fig:transition}), as portions of the VLA and VLBA are replaced by ngVLA items.

Multiple possible options exist for this VLA+VLBA to ngVLA transition.
At one extreme, the transition is gradual with the VLA and the VLBA kept operating at as-close-to-maximum scientific capability as long as possible as ngVLA capabilities are installed.
The benefit of this approach is that it maintains the overall scientific capabilities at centimeter wavelengths, at the potential cost of a substantially higher effective cost as the \hbox{VLA}, \hbox{VLBA}, and early ngVLA operations costs all must be supported. 
At another extreme, the transition could be as rapid as possible, essentially shutting down the VLA and VLBA entirely to be replaced by the \hbox{ngVLA}.
This option might be the most cost-effective, but it would imply a clear, unacceptable loss of scientific productivity and loss of scientific and technical capabilities due to the inability to use the VLA or VLBA or both.

In order to assess the various risks and benefits, the NRAO's ngVLA Project established the VLA+VLBA to ngVLA Transition Advisory Group (TAG).
This document presents the analysis undertaken by the TAG (Table~\ref{tab:tmatrix}) and its recommendations for the VLA+VLBA to ngVLA transition.  Specifically, \S\ref{sec:charge} presents the charge to the TAG and summarizes the TAG's evaluation process; \S\ref{sec:science} presents an analysis of the scientific opportunities of the next decade, focusing primarily on the \textit{Pathways to Discovery in Astronomy and Astrophysics for the 2020s} \citep{astro2020}, but drawing also upon \textit{Origins, Worlds, and Life: A Decadal Strategy for Planetary Science and Astrobiology 2023--2032} \citep{origin2023} and \textit{The Next Decade of Discovery in Solar and Space Physics: Exploring and Safeguarding Humanity's Home in Space} \citep{helio}; \S\ref{sec:technical} presents the technical options considered by the TAG; \S\ref{sec:find} presents findings and assessments of the technical options relative to their potential effects on the scientific opportunities; \S\ref{sec:recommend} describes the TAG's recommendations; and \S\ref{sec:engage} summarizes efforts undertaken by the TAG to encourage community engagement and feedback on the TAG's assessments and report.

\begin{table}[!b]
 \centering
\caption{{Decadal Science Priority Question-Transition Technical Option Assessment\label{tab:tmatrix}. Columns summarize science priority questions from the Decadal Survey report for which the VLA or VLBA or both are identified as ``very significant'' contributing instruments (\S\ref{sec:science}).  The rows summarize various technical options  (\S\ref{sec:technical}).  Each cell holds the assessment of the effect on the science priority question for implementing the corresponding technical option (\S\S\ref{sec:find}--\ref{sec:recommend}): \textbf{Not Acceptable (\hbox{N}, red)}---Science reduced very substantially and high-profile results likely inaccessible; \textbf{Moderately Acceptable (\hbox{M}, yellow)}---Science return reduced substantially, but acceptable for a limited duration and high-profile results likely at least partially accessible; \textbf{Acceptable (\hbox{A}, green)}---Science return is reduced, but high-profile results likely to remain accessible.}}
 \includegraphics[width=\textwidth]{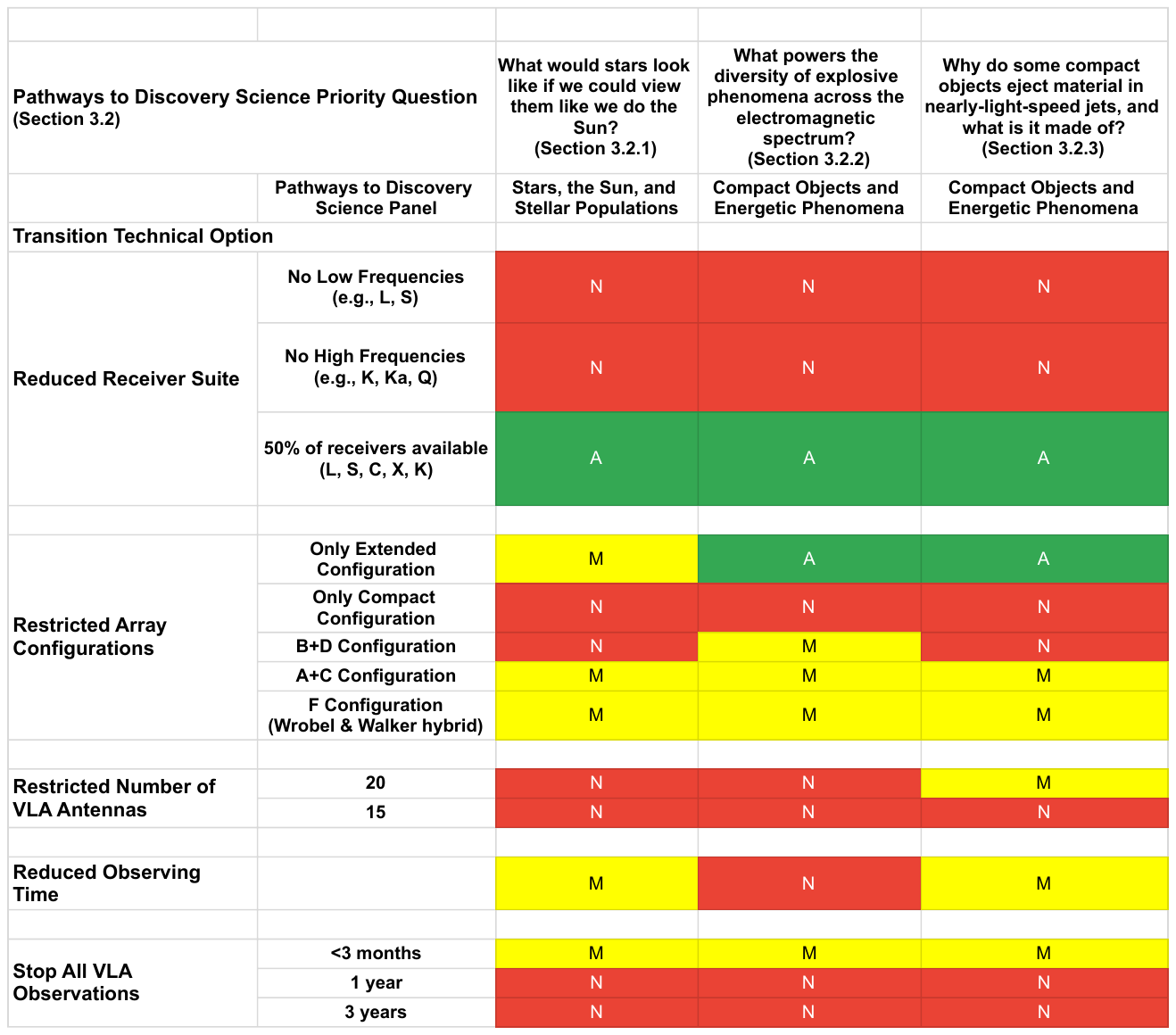}
\end{table}

\clearpage

\section{Charge and Evaluation Process}\label{sec:charge}

The VLA+VLBA to ngVLA TAG---a group of~18 members of the U.{}S.\ and international astronomical community---was charged to develop, quantitatively assess, and evaluate a set of possible VLA/VLBA to ngVLA transition options prioritized based on their scientific promise (given the scientific opportunities for the coming decade), of their cost, and their technical/personnel impacts.
Specifically, the TAG was tasked with:
\begin{itemize}
\item Identifying scientific opportunities in the coming decade that will critically benefit from complementary and/or unique observations at radio wavelengths.
\item Defining the relevant stakeholders affected by the transition (e.g., \hbox{NSF}, NRAO staff, astronomy community, etc.).
\item Identifying the relevant parameters of interest to critically compare transition options (e.g., VLA and VLBA observational capabilities, scientific areas impacted, archival science opportunities, staff load/sharing, etc.).
\item Identifying the necessary metrics by which to quantify the impact of different transition options.
\item Submitting a set of transition options to NRAO (i.e., the Internal Technical Analysis Team [ITAT]) for detailed (quantitative) costing/impact analysis.
\item Write up findings as part of the ``VLA/VLBA to ngVLA Transition Option Concepts'' report that includes a prioritized list of transition options.
\end{itemize}
The TAG main deliverable was defined as a ``VLA+VLBA to ngVLA Transition Option Concept'' report that would include a prioritized list of possible transition options, and intended to provide key input to NRAO and NSF planning.

In light of the above charge and expected deliverable, the TAG began its work by identifying and compiling VLA/VLBA/ngVLA science use cases, leveraging a variety of resources that included proposals submitted to the NRAO for observations with the VLA or \hbox{VLBA}, recent NRAO press releases, the \textit{ngVLA Science Book} \citep{ngVLAScienceBook}, and white papers submitted to the \textit{Pathways to Discovery} Decadal Survey \citep{astro2020}.
The TAG then constructed a ``matrix'' of science use cases (\S\ref{sec:science}) vs.\ specific technical transition options (\S\ref{sec:technical}) based on an initial set of technical options for the transition put forward by the \hbox{ITAT}.
The TAG further evaluated the scientific impact of each technical option via extensive bi-weekly discussions, and consulted the ITAT to assess the broad cost and personnel impact of each option (\S\ref{sec:find} and Table \ref{tab:tmatrix}). 
The TAG ultimately reached consensus on a transition plan that includes ranked options and foresees no reduction in VLBA capabilities during the VLA to ngVLA transition (\S\ref{sec:recommend}).

While conducting its work, the TAG advertised its activities broadly and received input from the scientific community (\S\S\ref{sec:other}--\ref{sec:engage}).
The TAG made this report available in a public forum (arXiv) and accept comments for up to 90~days before producing this final version.

\clearpage
%%%%%%%%%%%%%%%%%%%%%%%%%%%%%%%%%%%%%%%%%%%%%%%%%%%%%%%%%%%

\section{Scientific Opportunities During the VLA/VLBA to ngVLA Transition}\label{sec:science}

As a first and fundamental step in evaluating technical options, the TAG focused on identifying science use cases for a VLA+VLBA to ngVLA transition instrument, initially without considerations for which specific science might be of the highest opportunity.  
Perhaps not surprisingly, the science use cases identified by the TAG broadly mapped onto the priorities highlighted in \textit{Pathways to Discovery in Astronomy and Astrophysics for the 2020s} \citep{astro2020}, which charts an ambitious future for the next decade.
Both the main report and the report of the Panel on Radio, Millimeter, and Submillimeter Observations from the Ground \citep[Appendix~\hbox{M}, hereafter RMS panel]{astro2020} describe the broad contributions of radio wavelength observations to our current understanding of the Universe, and establish scientific priorities and opportunities for the next decade.

Without attempting to revisit such priorities, but rather adopting them as representative of the vision for the next decade from the broad scientific community, here we highlight areas of particular opportunity during the transition (\S\ref{sec:sc-1}--\S\ref{sec:sc-3}).
To this end, we discuss the Science Frontier Panel Questions/Discovery Areas identified in the ``High-Priority Science Questions Versus RMS Facilities'' \citep[Table~M.1]{astro2020} for which the \hbox{VLA}, the \hbox{VLBA}, or both are identified as ``very significant'' contributing instruments.
We stress that, in our discussion, we omit science cases for which the ngVLA would have an ``irreplacable and unique'' role (with no contribution from the VLA or VLBA) so as to keep our focus on the scientific opportunities of the transition era.

We also highlight scientific priorities identified in \textit{Origins, Worlds, and Life} \citep{origin2023}, released subsequently to \textit{Pathways to Discovery}, to which observations by the VLA or the VLBA could contribute today (\S\ref{sec:planetary}).
Finally, also subsequent to the release of \textit{Pathways to Discovery} and during the development of this report, the Decadal Survey for Solar and Space Physics (``Heliophysics'') was conducted, culminating in the release of \textit{The Next Decade of Discovery in Solar and Space Physics: Exploring and Safeguarding Humanity's Home in Space}; its recommendations as they relate to the VLA and VLBA are incorporated here (\S\ref{sec:helio}).

Throughout this summary, we illustrate recent VLA or VLBA contributions to various science areas, but we do not aim to provide a comprehensive review of the literature nor of all of the contributions of these telescopes.

\subsection{Discovery Areas Supported by the VLA/VLBA}
\label{sec:sc-2}

\textit{Pathways to Discovery} \citep{astro2020} highlighted several scientific discovery areas that are ripe for rapid progress in the coming decade.  Two of these areas, identified by the ``Compact Objects and Energetic Phenomena'' and ``Stars, the Sun, and Stellar Populations'' panels, would be supported by VLA and/or VLBA observations \citep[Tables~2.2 and~\S{M.1}]{astro2020}. 
We discuss the continuing and potential contributions of the VLA or VLBA to these discovery areas below. 

\subsubsection{Transforming our view of the universe by combining light, particles, and gravitational waves}
\label{sec:disc-2}
In recent years, time-domain and multi-messenger astronomy has moved to the forefront of modern astrophysics. Central to this development were the discovery of electromagnetic radiation associated with a gravitational wave signal from a binary neutron star merger, GW170817 %
\citep[e.g.,][and references within]{GW170817press,Abbott2017a,Abbott2017b}, and neutrinos associated with a flaring blazar~\cite{Aartsen2018}.
In both of these cases, there was a close time-coincidence with a flash of gamma rays, but radio observations provided crucial and unique information on the sources, in particular their relativistic outflows.
Broad-band radio light curves, as probed by the \hbox{VLA}, provide insight into the dynamics, structure, and evolution of jets launched in compact binary mergers \citep[e.g.,][and references within]{Alexander2017,Hallinan2017,Lazzati2018,Mooley2018a,Hovatta2021,Balasubramanian2022}.
In AGN jets, radio emission is a good proxy for the general jet activity and radio emission has been suggested to correlate with high-energy neutrino emission \citep[e.g.,][]{2020ApJ...894..101P,2021ApJ...908..157P,2024MNRAS.527L..26S}.
Very-long baseline interferometry allows for constraints on the expansion rate of the jet \citep{GW170817press2} and, in the case of the gravitational-wave event GW170817, it was the conclusive evidence that the jet was observed off axis \citep{Mooley2018b,Ghirlanda2019}. 

In the era of significant upgrades of gravitational wave and neutrino observatories, observations at radio wavelengths will remain key for making progress in these areas of research.
An emerging area is the TeV emission that is now being observed by Cherenkov telescopes of, for instance, gamma-ray bursts \citep{LHAASO2023}, challenging models of particle acceleration to the highest gamma-ray energies.
Also in this case, radio observations across various spectral, time, and spatial scales provide unique clues to, and context for, the origin of the highest-energy light and particles in the Universe.

\subsubsection{``Industrial Scale'' Spectroscopy}
\label{sec:disc-1}
The VLA and the VLBA have produced spectroscopic images naturally since their first operations.  With recent enhancements, the VLA in particular is producing spectrally-resolved images on an ``industrial scale,'' with images routinely having spectral coverage across fractional wavelength ranges or bandpasses of order unity.
The combination of increased sensitivity and spectral coverage and resolution has enabled the VLA to characterize stars across the Hertzsprung-Russell diagram, probing aspects of their quiescent emission, the frequency and energetics of their flaring activity, and the chemistry and physical conditions of their outflows and circumstellar environments \citep[e.g.,][]{2015ApJ...808..189W,2017A&A...599A.127F,2017A&A...602A..57A,2018A&A...612A..48W,2021AJ....161..111D,2023AJ....165...92M}.

Moreover, the VLA has been used to conduct high-cadence  spectroscopic imaging studies of solar flares, bursts, coronal mass ejections, and other energetic phenomena \citep[e.g.,][]{2021ApJ...911....4L}, allowing the Sun both to serve as a benchmark for other stars and to place it into a larger stellar context.

\subsection{Scientific Areas with VLA/VLBA Very Significant (and ngVLA Unique) Contribution}
\label{sec:sc-1}

\begin{figure}[tb]
    \centering
    \includegraphics[width=0.95\textwidth]{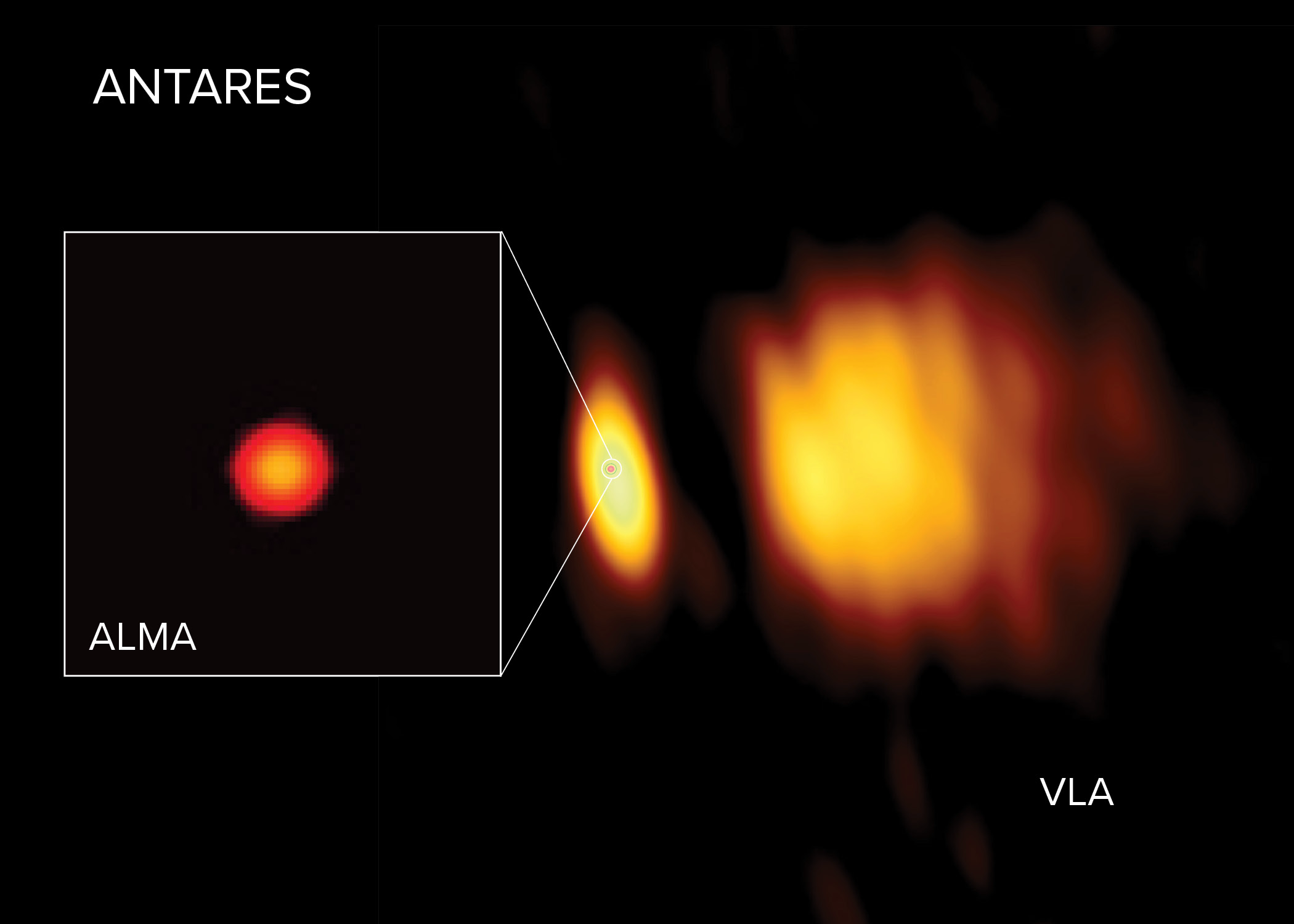}
    \caption{The frequency coverage and angular resolution of the VLA enables it to begin developing three-dimensional views of other stars in a manner analogous to how we can view the Sun, such as Antares as illustrated here.  In this image, not only is Antares itself resolved, a portion of its wind also is visible on the right, illuminated by its companion Antares~\hbox{B}.  VLA observations can be augmented by observations at higher frequencies/shorter wavelengths, such as ALMA illustrated here, to probe deeper into stellar atmospheres via spectroscopic observations.  Figure credits: ALMA (ESO/NAOJ/NRAO), E.~O’Gorman; \hbox{NRAO/AUI/NSF}, S.~Dagnello.}
    \label{fig:Stellar}
\end{figure}

\begin{figure}[bt]
    \centering
    \includegraphics[width=\textwidth]{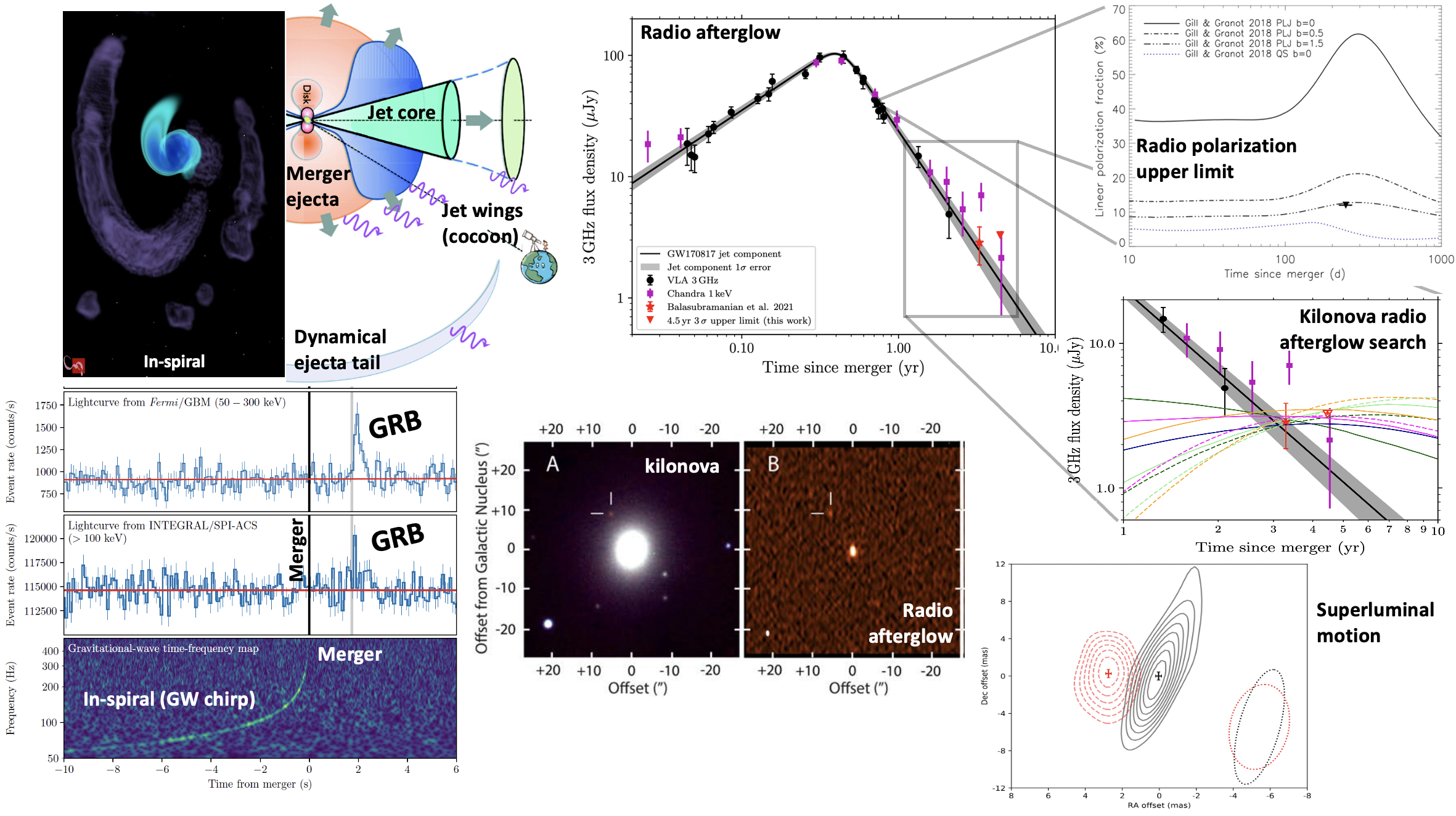}
    \caption{Over the next decade, the VLA and the VLBA will provide crucial observations contributing to the multi-messenger understanding of explosive events and relativistic jets, thereby addressing multiple science priority questions.  An exemplary case was that of \hbox{GW~170817/GRB~170817A}, for which the VLA observations of the radio afterglow (middle panels; \citep{Hallinan2017,2022ApJ...938...12B} and references therein) informed both the calorimetry and structure of the ejecta.  In the case of the ejecta structure, both the total intensity and the polarization measurements place strong constraints on the nature of the (structured) jet that emerged \citep{Mooley2018a,Mooley2018b,2018ApJ...861L..10C,2022ApJ...938...12B}.  The VLBA observations (right panel) demonstrate convincingly that this event produced a superluminal jet \citep{mooley2018}.  VLA and VLBA observations of similar phenomena will contribute in an analogous manner during future ground-based gravitational wave observatory runs, and they may contribute to such analyses associated with neutrino events from observatories such as IceCube or IceCube-Gen2. Figure adapted from \citep{Abbott2017a,Hallinan2017,Ioka2018,2022ApJ...938...12B,2018ApJ...861...12L,Mooley2018b}. Additional Figure credits: NRAO/AUI/Max Planck Institute for Gravitational Physics.}
    \label{fig:Explosive}
\end{figure}

Three science priority questions were identified as those to which observations by the VLA and/or VLBA today ``would make a very significant contribution in addressing [...] but would not be sufficient to address that question by itself (e.g., in the absence of observations at other wavelengths)'' (Figures~\ref{fig:Stellar} and~\ref{fig:Explosive}).
Going forward, the ngVLA is expected to address these science priority questions at different levels. Specifically, the ngVLA will be essential but not sufficient by itself for addressing the question ``What would stars look like if we could view them like we do the Sun?'' (\S\ref{sec:stars}, Figure~\ref{fig:Stellar}).
The ngVLA would play a role that is ``irreplaceable and unique relative to other facilities with U.{}S.\ community access'' for addressing the questions (Figure~\ref{fig:Explosive}) ``What powers the diversity of explosive phenomena across the electromagnetic spectrum?'' (\S\ref{sec:explosions}) and ``Why do some compact objects eject material in nearly-light-speed jets, and what is it made of?'' (\S\ref{sec:jets}).

\subsubsection{What would stars look like if we could view them like we do the Sun?} 
\label{sec:stars}

The VLBA enables imaging the structures around evolved stars \citep[e.g.,][]{2013MNRAS.433.3133G,2018ApJ...853...42S}, which reflect how these stars lose mass and often do so asymmetrically.  In turn, such observations both constrain models for mass loss and foreshadow the end states of stars, either as planetary nebulae for low-mass stars or the potential interactions between supernovae shocks and the surrounding circumstellar media for high-mass stars.

Imaging of thermal emission from radio photospheres and chromospheres can obtain the direct measurements of the diameters of stars \citep[e.g.,][]{2012A&A...544A..42Z}.  Of particular promise are combined millimeter- (Atacama Large Millimeter/submillimeter Array [ALMA]) and centimeter-wave (VLA) observations, which have provided among the most detailed radio map yet of any star, other than the Sun \citep{SupergiantPress,2020A&A...638A..65O}.  Combining VLA and near-infrared wavelength observations can measure wind speeds on brown dwarfs 
\citep{BrownDwarfPress,2020Sci...368..169A}, 
likely resulting from zonal wind patterns, and potentially analogous flows in (very) low-mass stars.
Astrometric observations conducted with the VLBA can reveal the presence of lower-mass companions \citep[e.g.,][]{VLBAPlanetPress,VLBAPlanetPress2,2016ApJ...827...22F,2016ApJ...827...23D,Curiel_2020,2024ApJ...967..112C}, which provide a more complete description of stellar systems and can assist with discriminating between stellar surface features and companions.

\subsubsection{What powers the diversity of explosive phenomena across the electromagnetic spectrum?}
\label{sec:explosions}

Observations by the VLA and the VLBA across their entire operational frequency range provide critical information about the existence of fast ejecta (from supersonic ejecta to highly-relativistic jets) in a variety of astrophysical sources that include stellar eruptions \citep[e.g.,][]{2021ApJ...911....4L}, stellar explosions \citep[e.g.,][]{2020ApJ...892...82R}, stellar disruptions by supermassive black holes \citep[e.g.,][]{2021ARA&A..59...21G}, mergers of stars \citep[e.g.,][see also Figure \ref{fig:Explosive}]{SupernovaPress,GW170817press2,Hallinan2017,Mooley2018a,Mooley2018b,mooley2018,2022ApJ...924...16M}, fast radio bursts \citep[e.g.,][]{FRBPress,2019ApJ...886...24L,2020ApJ...899..161L,2022Natur.606..873N}, and potentially discoveries of new classes of sources \citep[e.g.,][]{CBlastPress,CExplosionPress,2020ApJ...895L..23C,2020ApJ...895...49H}.

These observations, including surveys of the radio sky \citep[e.g.][]{VLASSPress,2018ApJ...866L..22L} and long-term monitoring observations, also provide information on the processes behind relativistic particle acceleration and magnetic field amplification,  nature of the interactions between these ejecta and the ambient media, on the mass-loss history of massive stars before core collapse, on the synthesis of the heaviest elements of the periodic table, and on the intergalactic medium.

Particularly in the context of follow-up observations of gravitational-wave \citep{GW170817press,GW170817press2} and high-energy neutrino events
\citep{NeutrinoPress,2018Sci...361.1378I}, observations by the VLA or the VLBA or both provide crucial multi-messenger information about the nature of these events (\S\ref{sec:disc-2}).
     
\subsubsection{Why do some compact objects eject material in nearly-light-speed jets, and what is it made of?} 
\label{sec:jets}
The physics of jets has been intensively studied for many years with a variety of observational
and theoretical techniques and in the context of a suite of astrophysical sources including supermassive black holes (SMBHs) in AGN, stellar-mass black holes and neutron stars in X-ray binaries, gamma-ray bursts, stripped-envelope core-collapse supernovae, and tidal disruption
events \citep{2019Natur.569..374M,2021ApJ...923L...5P,2021ApJ...911L..11E,2021ApJ...909...80B,2021AJ....161..207M,2021AN....342.1117L}. However, these jets  are still poorly understood. Both the VLA and VLBA have a laudable history of studying highly relativistic jets, as illustrated by a significant number of press releases in the past decade \citep{JetPress1,JetPress2,JetPress3,JetPress4,JetPress5,JetPress6,JetPress7}. These observations have acquired a new urgency in the context of both multi-messenger science (Figure \ref{fig:Explosive}) and improved computational modeling efforts.  Observations across the entire operational frequency range of the VLA and/or VLBA trace the relativistic material in the jet from where it launches to where it interacts with the ambient medium, which may be tens of kiloparsecs from the compact object. Radio observations (including polarization measurements) are also critical to help determine the jet composition and magnetic field structure \citep[e.g.,][and references therein]{1999MNRAS.309L...7G,2018MNRAS.478.4128G,2018ApJ...861L..10C}.

\subsection{VLA/VLBA Support to Other Scientific Areas (and ngVLA Irreplaceable Role)}
\label{sec:sc-3}

The RMS panel identified a variety of other scientific questions that today's observations by the VLA or VLBA or both would ``have an impact in addressing [...], but would be one of several facilities playing supporting roles.''
In what follows, we list some of these science questions in the order that the RMS Panel had them. 
We stress that, with respect to the questions described in \S\S\ref{sec:sc-3.q1}--\ref{sec:sc-3.q3} and \S\ref{sec-q10} below, the ngVLA would play a role that is ``irreplaceable and unique relative to other facilities with U.{}S.\ community access.'' 

\subsubsection{How do star-forming structures arise from, and interact with, the diffuse ISM? }\label{sec:sc-3.q1}
    
The VLA and VLBA are playing important roles in supporting the suite of multi-wavelength observations that are mapping in unprecedented detail the 3-D structure of the dust and protostars in star-forming regions, probing star-forming structures and their motions, and tracking the effects of stellar feedback \citep{MWPress,NurseryPress,DustPress}.
Observations with the VLA can be used to trace the diffuse gas (21~cm), the raw material for all star formation, and a variety of spectral lines that trace both molecular gas and the ionized gas that results from star forming regions forming from and interacting with their environments \citep[e.g.,][]{2013MNRAS.429..987L,2020AJ....160..234D,2020ApJ...896L...3D,2021A&A...651A..85B}.
Observations with the VLBA can use masers as indicators of high-mass star formation \citep[e.g.,][]{2019ApJS..244...35L}, place constraints on the magnetic field, and measure the distances to younger systems \citep[e.g.,][]{2021ApJ...906...24D}.
These VLA and VLBA observations can be used in the context of existing and forthcoming surveys, such as \textit{Herschel}, Gaia, the Vera Rubin Observatory's Legacy Survey of Space and Time (LSST), the Sloan Digital Sky Survey (SDSS)~\hbox{V}, and the Spectro-Photometer for the History of the Universe and Ices Explorer (SPHEREx), and they complement the pointed observations being conducted and planned for \textit{JWST} and \hbox{ALMA}.

\subsubsection{What regulates the structures and motions within molecular clouds?}\label{sec:sc-3.q2}
    
Observations with the VLA provide critical information about the magnetic field and the physical conditions (density, and temperature) within molecular clouds, and astrometric observations with both the VLA and VLBA over the course of several years to a decade or more can track the motions of stars and outflows \citep[e.g.,][]{2017ApJ...834..139D,2021ApJ...906...24D}.
Numerical simulations suggest that molecular clouds are formed when flows within the diffuse interstellar medium collide \citep[e.g.,][]{2014ApJ...791..124G}, and gravitational collapse starts rapidly upon cloud formation and results in the emergence of filamentary structures and dense cores embedded within the filaments.  These expectations are borne out by a number of observations of molecular clouds, most notably with \textit{Herschel}.  Subsequently, molecular clouds are shaped by competing physical processes ranging from external flows caused by stellar feedback (e.g., expanding H\,\textsc{ii} regions, supernova blast waves) to magnetic fields to energy and momentum injected by protostellar outflows.  Magnetic field information within these molecular cloud filaments can be obtained from VLA observations of the Zeeman effect in OH molecules, while observations of the density- and temperature-sensitive molecule ammonia can trace infall motions and velocity oscillations along filaments that lead to core formation for a variety of gas densities.  These VLA and VLBA observations complement those of ALMA by probing different condensations within the molecular gas.  Stars born from the densest regions of molecular clouds inherit the dynamics of the parental clouds and filaments, resulting in typical velocities of several kilometers per second, which are easily detectable through VLA and VLBA astrometric observations.

\subsubsection{How does gas flow from parsec scales down to protostars and disks?}
\label{sec:sc-3.q3}

From scales of parsecs to of order 1000~au, VLA observations provide resolved measurements of the density structures of the dense cores within which stars are forming \citep[e.g.,][]{2012ApJ...748...16T,2019ApJ...882..103P,2020A&A...644A.128S,2020ApJ...890..129K,2020ApJ...890..130T,2022MNRAS.516..185K}, thereby tracking the processes occurring from the parsec-scale turbulent flows to accreting protostellar and protoplanetary disks \citep{PlanetBirthPress}.
Among the objectives is to understand how mass is transported inward and angular momentum is transported outward, resulting in the apparently universal stellar initial mass function (IMF) and planet-forming disks.  These observations include the (polarized) dust continuum emission, the velocities and line widths of spectral lines of key molecules such as ammonia, and Zeeman splitting of spectral lines.  With these, the internal kinematics \citep[e.g.,][]{2012ApJ...746..174R}, density fluctuations, and magnetic field structures \citep[e.g.,][]{2022MNRAS.516L..48K} can be inferred, thereby constraining the lifetimes of cores, their susceptibilities to fragmentation and binary star formation, and linking core masses to the stellar IMF.  VLA observations, when combined with those from ALMA and potentially future far-IR missions, are key to penetrating the high optical depths present in high-mass star formation in environments with very high dust extinction, and the densest parts (generally smallest-scales) of infalling low-mass cores.

\subsubsection{How do the Sun and other stars create space weather?}\label{sec:sc-3.q4}
    
Multi-frequency observations with the VLA and the VLBA of stars of different spectral types help place the Sun’s mass loss and interactions with its environment, a.k.a. ``space weather,'' into a broader context.  High spatial and temporal resolution observations with the VLA allow the temporal and spatial evolution of stellar radio bursts to be followed \citep[e.g.,][]{2018ApJ...856...39C,2019ApJ...871..214V}.
Likely driven by magnetic activity similar to that that occurs on the Sun, such VLA observations contribute to constraining the properties of the coronal magnetic fields on other stars and the potential production of non-thermal particles that can affect the habitability of extra-solar planets.
Particularly for evolved stars, VLA and VLBA observations are not limited by dust obscuration, and they can be used to characterize and track stellar ejecta \citep[e.g.,][]{2020ApJ...901...98L}.
Characterizing such stellar ejecta both constrains the processes by which the stellar mass loss occurs and provides a full measure of the current mass loss.  In synergy with mid-IR observations, the total mass (gas + dust) of the ejecta and its history can be reconstructed, which is relevant in the context of the chemical enrichment of the \hbox{ISM}.

Finally, monitoring of the non-thermal radio emission from stars, over a wide frequency range, has the potential to reveal signatures of ``extrasolar space weather,'' such as coronal mass ejections from other stars \citep{2017PhDT........85V}.  In addition to providing ``extrasolar space weather'' events against which to compare and contrast those from the Sun, the detection and characterization of such ``extrasolar space weather'' events may be critical for assessing the potential habitability of planets.

\subsubsection{What are the mass and spin distributions of neutron stars and stellar mass black holes?}\label{sec:sc-3.q5}

The VLA observes millisecond pulsars as part of the North American Nanohertz Observatory for Gravitational Waves 
\citep[NANOGrav,][]{2023ApJ...951L...9A}.
Obtaining the precision pulsar timing required for gravitational wave observations places stringent requirements on other potential contributions, such as changes in the radio pulse time of arrival for those pulsars in binary systems.
Consequently, the same precision pulse timing observations can be used to measure the characteristics of radio pulsars in binary systems, such as determining relativistic Shapiro delays, which in turn lead to stringent measurements of neutron star masses, including the most massive neutron star known to date \citep{cromartie2020}. 
Relative to many other telescopes, the VLA provides both increased frequency coverage and increased bandwidth, which are useful to mitigate interstellar propagation effects that otherwise would vitiate such measurements
\citep[e.g.,][]{2018ApJ...861...12L}.

\subsubsection{What seeds supermassive black holes and how do they grow?}\label{sec:sc-3.q6}

The VLA and the VLBA can be used to search for black holes in
or near the centers of dwarf galaxies \citep[e.g.,][]{WanderingBHPress,2020ApJ...888...36R,2022ApJ...941...43Y,2022ApJ...933..160S}, which are expected to have masses in the regime of intermediate mass black holes (IMBHs, $10^2\,\mathrm{M}_\odot$--$10^5\,\mathrm{M}_\odot$; \citep{2020ARA&A..58..257G}).  IMBHs represent one of the potential seeds for SMBHs, most notably for those SMBHs found at high redshift ($z > 5$) for which growth from stellar masses ($\sim 10\,\mathrm{M}_\odot$ is challenging in the available time at high redshifts.  The VLA and VLBA observations help constrain the local population of such SMBH seeds, which in turn constrains the mechanisms at high redshift.
The VLBA also can be used to search for pairs of active galactic nuclei (\hbox{AGN}, e.g., \citep{2023ApJ...958...29C}), and the VLBA is the only telescope with the resolution capable of detecting pairs with parsec-scale separations.  At these scales, pairs of SMBHs likely form bound binaries, and potentially enter the regime of emitting gravitational waves, at which which point they are destined to merge.
Coupled with multi-wavelength and multi-messenger observations from the full suite of astronomical facilities, the VLA and VLBA will contribute to constraining the extent to which IMBHs might have formed the seeds for the current population of SMBHs and the extent to which SMBHs grow by mergers rather than by accretion.
    
\subsubsection{How did the intergalactic medium and the first sources of radiation evolve from cosmic dawn through the epoch of reionization?} 
    
The principal source of reionizing photons in the early Universe is thought to be a population of low luminosity, low mass star forming galaxies. However the number density of active supermassive black holes (quasars, radio galaxies) is presently uncertain. There are now numerous confirmed detections of quasars within the cosmic reionization epoch \citep[e.g.][]{wang2021}, and although these objects are thought to be a sub-dominant source of ionizing photons, they can act as signposts for over-densities of galaxies around the most massive dark matter haloes \citep{endsley2022}.

Due to their angular resolution and multi-band receiver suites, the VLA and VLBA play a crucial role in the identification and characterisation of high redshift active black holes, as the radio emission from such systems is compact and exhibits a steep spectrum due to the inverse Compton processes associated with the higher temperature cosmic microwave background. Radio luminosities in excess of what would be expected from purely star-formation driven synchrotron are also a powerful diagnostic for activity associated with accretion onto a supermassive black hole.

Additionally, the VLA is presently the by far most sensitive instrument capable of targeting the lower $J$ transitions of the $^{12}$CO molecule at redshifts approaching and into the cosmic reionization epoch, thereby providing the best estimates of the total molecular gas budget for star formation. It should be noted however that it is challenging to obtain such detections in all but the most extreme systems.

\subsubsection{How do gas, metals, and dust flow into, through, and out of galaxies?}

The \hbox{VLA}, and to some extent the \hbox{VLBA}, can trace atomic and molecular gas and cosmic rays flowing into and being ejected from galaxies.  Observing with the distinctive 21\,cm line of H\,\textsc{i}, the VLA can characterize the large-scale distribution and kinematics of the diffuse gas in and around nearby galaxies and groups, either pristine gas inflowing or gas that has been stripped by galaxy interactions \citep[e.g.,][]{2017AJ....153..132A,2018MNRAS.473.5248O,2023A&A...670A..21J}.
Both the VLA and the VLBA can trace radio jets, which may be responsible for driving outflows, across a range of spatial scales (as described elsewhere in this report).
Moreover, both the VLA and the VLBA can characterize outflows or the interactions of outflows with ambient material via absorption and emission studies \citep[e.g.,][]{2013ApJ...779..173N,2018A&A...617A..38S}.

Crucially, observations at radio wavelengths, such as those by the \hbox{VLA}, provide one of the few means of tracking and characterizing non-thermal processes, namely cosmic rays and magnetic fields, that could drive galactic winds and more generally affect galaxies, their circumgalactic media, and the larger intragroup or intercluster media \citep[e.g.,][]{CosmicRayPress,RopePress,2015AJ....150...81W,2019MNRAS.487.1498W,2022MNRAS.517.2990T,2024Galax..12...22I}.

These VLA and VLBA form part of the multiwavelength characterization of the baryon cycle through galaxies, complementing molecular line observations by ALMA or the Northern Extended Millimeter Array (\hbox{NOEMA)}, observations of both neutral and ionized species by current and future telescopes operating from the infrared to the ultraviolet, X-ray observations of hot gas by \textit{Chandra} and potentially future X-ray telescopes, and potentially even $\gamma$-ray observations by \textit{Fermi}.

\subsubsection{How do supermassive black holes form and how is their growth coupled to the evolution of their host galaxies?}

The VLA and VLBA have played a leading role in the study of SMBH feedback, by tracing the synchrotron radio emission from jetted AGN \citep{DoubleHelixPress,StormPress,BlowingPress} and probing the amount of energy transferred to interstellar media or intracluster media in conjunction with complementary multi-wavelength data (e.g., from X-ray telescopes) \citep{QuartetPress}.

As described elsewhere in this section, the VLA and the VLBA play crucial roles in tracking material potentially flowing toward SMBHs or AGN-driven outflows from galaxies, which could regulate the star formation of their host galaxies
and the effect of SMBH feedback on even larger scales \citep[e.g.,][]{WeakBHPress,2019MNRAS.484.3376L,2023A&A...670A..23U}.

These VLA and VLBA observations are part of a multi-wavelength approach to characterizing SMBH growth and feedback, complementing molecular line observations by ALMA or the Northern Extended Millimeter Array (\hbox{NOEMA}), observations of both neutral and ionized species by current and future telescopes operating from the infrared to the ultraviolet, and X-ray observations of hot gas by \textit{Chandra}.

\subsubsection{How do the histories of galaxies and their dark matter halos shape their observable properties? }\label{sec-q10}

From the discovery of spectacular ``threads'' and ``filaments'' \citep{1984Natur.310..557Y,1985AJ.....90.2511M} to on-going work to trace star formation and the density and temperature structure of star-forming regions 
\citep{2019MNRAS.482.5349R,2022ApJ...936..186B,2022A&A...666A..31M,2023A&A...680A..43L}, the VLA continues to provide new insights into the extreme environment of the Milky Way's Galactic center.
Serving as the local model for high-pressure environments that might occur in other, more distant galaxies, the Galactic center is a region in which a variety of different environmental effects (turbulence, magnetic fields, cosmic rays, \ldots) can be investigated for their effects on the efficiency of star formation \citep{2021NewAR..9301630B}.  
Because radio wavelengths suffer essentially no extinction toward the Galactic center, the VLA will continue to be a crucial element in characterizing the properties of the Galactic center over the next decade.

\subsubsection{What are the properties of dark matter and the dark sector?}

Interpretations of results, or projections of future results, from the NANOGrav include the possibility that nanohertz gravitational waves could reveal characteristics of dark matter within the Galaxy \citep{2023ApJ...951L..11A}.  The VLA contributed to some of the NANOGrav observations, and it could continue to do so, particularly in light of the loss of the Arecibo Observatory. Measurements of binary pulsars accelerations constrain the local dark matter density \citep{chakrabarti2021}. The VLA also may be able to contribute observations that would help reveal whether the ``Galactic center excess'' discovered in data obtained by the \textit{Fermi} Gamma-Ray Telescope is due to emission from millisecond pulsars or results from decay products from dark matter particles \citep{2017MNRAS.468.2526B}.

\subsubsection{How will measurements of gravitational waves reshape our cosmological view?} 

Recent interpretations of results from the NANOGrav have included the possibility that nanohertz gravitational waves could result from physics in the early Universe beyond the standard hot Big Bang model \citep{2023ApJ...951L...8A,2023ApJ...951L..11A}.  The VLA contributed to some of the NANOGrav observations, and it could continue to do so, particularly in light of the loss of the Arecibo Observatory.

Relevant to higher gravitational wave frequencies, VLBA+VLA observations of the superluminal motion of the jet from GW~170817 \citep{mooley2018}, combined with previous gravitational wave and electromagnetic data, led to an independent constraint on Hubble's constant \citep{hotokezaka2019}. Additional localizations can provide constraints on the Hubble constant projected to be comparable to current CMB and Cepheid-supernova observations.

\subsection{VLA and VLBA Contributions to Solar System and Astrobiology Science Questions }
\label{sec:planetary}

While the RMS Panel identified no science priority question in the area of extrasolar planets, astrobiology, or the Solar System to which observations by the VLA or the VLBA could contribute today, \textit{Origins, Worlds, and Life} Decadal \citep{origin2023} was released subsequently to \textit{Pathways to Discovery}.  There are several high priority Solar System and astrobiology science questions that can be answered in part by the \hbox{VLA}, including the ones described below.

In addition to the specific science questions discussed below, there are broader contributions from the \hbox{VLA}, \hbox{VLBA}, or both to the field of Planetary Science and Astrobiology.  
\begin{itemize}
    \item%
    Astrometric observations with the VLBA have contributed to improved orbits of Solar System planets, providing improved navigation capabilities.
    \item%
    The \textit{Origins, Worlds, and Life} report recommended that NSF and NASA conduct a study to assess the needed capabilities for ground-based planetary radar observations, particularly in light of the collapse of the Arecibo Observatory.  The resulting study, ``Cross-Disciplinary Deep Space Radar Needs Study,'' is now available \citep{SpaceRadar},
    and both the VLA and VLBA are identified as existing receiving facilities for the Goldstone Solar System Radar and for a potential planetary radar capability at the Green Bank Telescope (GBT).  The role of the VLBA in proof-of-concept planetary radar observations with the GBT has been highlighted \citep{DefensePress,TychoPress}.
\end{itemize}

\subsubsection{Q1: Evolution of the Protoplanetary Disk}\label{sec:sc-4:q1}

The longer wavelengths at which the VLA can observe probe the protoplanetary disk regions that are optically thick in the shorter-wavelength ALMA observations \citep[e.g.,][]{2016A&A...588A.112G,2016ApJ...821L..16C,2019ApJ...883...71C}, revealing details of their substructures as well as placing constraints on dust growth in the disk. These observations are critical for placing disk evolution in context with the protostellar properties.

\subsubsection{Q2: Accretion in the Outer Solar System}\label{sec:sc-4:q2}

Radio observations with the VLA have been, and continue to be, instrumental in understanding the bulk abundances of the giant planets \citep[][provides a review]{depater23}, with implications for Q2.1: How did the giant planets form? and Q2.2: What controlled the compositions of the material that formed the giant planets?

\subsubsection{Q7: Giant Planet Structure and Evolution}\label{sec:sc-4:q7}

The VLA is uniquely capable of spatially resolving the giant planets at depths up to about~50\,bar pressures, providing a window into the deep temperature structure,
chemical abundances, and dynamical processes in these atmospheres that is completely inaccessible at other wavelengths \citep[e.g.,][]{JupyterPress,2019Icar..322..168D,2022PhDT........15M,2023PSJ.....4...25M,2023GeoRL..5002872A}.
Moreover, with the \hbox{VLA}, observations can be obtained over much longer durations, providing important context for spacecraft missions of shorter duration, such as Juno, the Jupiter Icy Moons Explorer (Juice), and the Europa Clipper.

\subsubsection{Q8: Circumplanetary Systems}\label{sec:sc-4:q8}

Multi-wavelength observations of ring systems with the VLA can obtain key information about ring particle properties such as porosity and about their potential interactions with planetary magnetospheres or satellites.  In the past decade, much of the focus has been on Saturn's ring system \citep{2017PhDT.........3Z,2019Icar..317..518Z}, largely because of the Grand Finale phase of the \textit{Cassini} mission.  In light of the recommended Uranus Orbiter \& Probe mission, and the importance of characterizing ring properties as part of that mission's design of its orbits, observations with the VLA may provide crucial information about the Uranian ring system.

With the launch of the Dragonfly mission scheduled for later this decade, VLA observations of Titan could serve as a key element of setting the context for that mission, particularly in concert with likely ALMA observations.

\subsubsection{Q12: Exoplanets}\label{sec:sc-4:q12}

The observations discussed in the previous bullets also carry relevance to exoplanets; comparisons between the present-day state of the outer Solar System and the observed properties of exoplanets and protoplanetary disks were identified as high-priority by the \textit{Origins, Worlds, and Life} report.

\subsection{VLA and VLBA Contributions to Solar and Space Physics (Heliophysics) Science Questions}
\label{sec:helio}

The RMS panel identified the science frontier question ``How do the Sun and other stars create space weather?'' as one to which the VLA could provide a supporting role in addressing.
\textit{The Next Decade of Discovery in Solar and Space Physics: Exploring and Safeguarding Humanity's Home in Space} Decadal survey report \citep{helio} was released subsequently to \textit{Pathways to Discovery}.
That report identifies the VLA as one of the major ground-based facilities developed by the \hbox{NSF} and used for solar and space physics investigations.
Moreover, the supporting report from the Panel on the Physics of the Sun and Heliosphere  highlights that one of the technical advances that has led to scientific discoveries has been ``the transition from interferometric imaging at a few discrete frequencies to true radio imaging spectroscopy over broad frequency bands'' \citep[e.g.,][]{2013ApJ...763L..21C}, echoing the ``Industrial-Scale Spectroscopy'' discovery area identified in \textit{Pathways to Discovery}.

\textit{Exploring and Safeguarding Humanity's Home in Space} identifies three Science Themes, two of which are ``A Laboratory in Space: Building Blocks of Understanding'' and ``New Environments: Exploring Our Cosmic Neighborhood and Beyond'' and can be addressed by current and future VLA observations.
For example, joint VLA and X-ray spectroscopic observations of solar flares responds to the Guiding Question ``How do fundamental processes create and dissipate explosive phenomena across the heliosphere?'' under the ``A Laboratory in Space'' Science Theme \citep{2022ApJ...940..137L}. 
There have been multiple VLA observations of other stars, often low-mass stars, aimed at understanding both how their magnetic field structures differ from that of the Sun and what the potential implications are for the habitability of planets around those stars \citep[e.g.,][]{2018ApJ...856...39C,2018ApJ...857..133B,2018ApJ...866..155P}.
These observations respond to the Guiding Questions ``Why Does the Sun and Its Environment Differ from Other Similar Stars?'' and ``What Internal and External Characteristics Have Played a Role in Creating a Space Environment Conducive to Life?'' under the ``New Environments'' Science Theme.  
Further, the Panel on the Physics of the Sun and Heliosphere discusses how knowledge of the interplanetary magnetic field is part of understanding ``dynamic solar processes'' throughout the heliosphere, with one of the figures supporting the discussion being from an analysis of measurements acquired with one of the instruments on Solar Terrestrial Relations Observatory (STEREO)-A and VLA observations \citep{2020ApJ...896...99W}.

\clearpage
\section{Technical Options for the Transition} \label{sec:technical} 

The TAG received from the NRAO ITAT a document entitled ``Transition Supporting Information'' outlining a set of technical options that could be considered in implementing the VLA+VLBA to ngVLA transition.
In general, these technical options involve reducing some measure of capability for either the VLA or the VLBA or both, with the aim of enabling NRAO to redirect staff efforts from VLA or VLBA operations to ngVLA construction. Further, the transition options were limited to hardware capabilities and did not encompass other critical aspects of the VLA and \hbox{VLBA}, such as the implementation of scientific operations, data processing, and data products delivery to the community.

The TAG consolidated these technical options into the following broad set, listed in no particular order:
\begin{description}
\item[Reduced Receiver Suite:]%
  The number of receivers maintained at the antennas could be reduced, reducing the set of frequency bands that could be observed.  

\item[Restricted Array Configurations:]%
  The number of configurations into which antennas are moved could be reduced or even restricted to a single configuration, reducing the range of angular resolution that the VLA could access, the range of surface brightness sensitivities that the VLA could probe, or both.

\item[Restricted Correlator Setups:]%
  The number of possible correlator modes supported could be reduced, reducing the flexibility of the VLA and decreasing the chances of new discoveries.

\item[Reduced Observing Time (no daytime observations):]%
  The amount of observing time available could be reduced, recognizing that observations during the daytime may have to be curtailed or ceased completely during the Transition due to safety considerations.

\item[Stop All VLA Observations:]%
  All VLA observations could be ceased for some limited duration, with the Transition Advisory Group considering a range from three months to three years.		
\end{description}

\smallskip
\noindent%
There also likely are a number of intermediate or ``hybrid'' approaches.
For instance, rather than retain the current full VLA receiver suite or shut down some of the receivers, the VLA might be able to be used as a testbed for ngVLA-like receivers.
There have been multiple designs published for wideband receivers \citep[e.g.,][]{8879392,2023IJMw....3..570K}, and one of these designs could be tested on the VLA antennas, allowing NRAO staff to begin to transition to maintaining ngVLA-like equipment while maintaining essentially full VLA capabilities.
Alternately, there could be an ``early deployment'' of ngVLA correlation or data processing, shifting such capabilities rapidly from the current VLA or VLBA to the \hbox{ngVLA}.
This latter approach may reduce the risk that current VLA or VLBA capabilities are increasingly difficult to maintain due to obsolescence of digital signal processing or computational equipment. 
The TAG did not consider such possibilities at length.

\clearpage
\section{Findings and Assessments}\label{sec:find}

TAG members provided a quantitative assessment of the effect of each
of the technical transition options (\S\ref{sec:technical}) on each of
the ngVLA science cases, which then were used as the basis for
assessing against the priority science questions identified in the
\textit{Pathways to Discovery} (\S\ref{sec:science}).

After this initial quantitative assessment, the results were discussed extensively during bi-weekly TAG meetings, and augmented with anonymous surveys to enable TAG members to provide additional unbiased assessments as the transition options were progressively shaped.

\subsection{Findings}\label{sec:find.find}

During the course of its review of material and deliberations, the TAG made the following scientific determinations.  These are listed below in no particular priority order.

\paragraph{The focus of the transition is on the VLA capabilities.}\label{par:vlba}%
Compared to the \hbox{VLA}, the VLBA has fewer antennas and they are immobile.  The consequence is that the opportunities to reduce the capabilities of the VLBA are both more limited and more likely to have a greater reduction in scientific return.  Further, there may be other considerations associated with reductions in the capabilities of the VLBA (\S\ref{sec:other}).

\paragraph{The frequency range of the VLA and VLBA are unique and provide compelling scientific capabilities.}\label{par:freq}%
The VLA can observe from~74~MHz to~50~GHz, and the VLBA can observe from~0.33~GHz to~86~GHz.  In both cases, the spectral dynamic range exceeds 100:1, unequaled by any current or near-term radio-millimeter-submillimeter facility.
These large spectral ranges enable observing a diverse range of sources, with a variety of emission mechanisms to be studied, and it allows the evolution of a source's emission in frequency to be tracked.

\paragraph{Time domain science is a compelling opportunity for the next decade.}\label{par:tdamm}
The exquisite combination of resolution and sensitivity of the VLA/VLBA, together with their PI-driven operation mode, have deeply impacted the field of time-domain and multi-messenger astronomy. Over the next decade, the ngVLA will bring multi-messenger time-domain astronomy to its full potential in combination with current and planned telescopes, particle detectors, and gravitational-wave observatories.

\paragraph{Near-term planetary science missions offer compelling opportunities for complementary observations.}\label{par:missions}
The VLA is uniquely capable of providing the high spatial resolution, wide bandwidth, and dense $u$-$v$ plane coverage necessary for radio observations of Solar System planets. These are especially valuable when combined with spacecraft observations; the Jupiter Icy Moons Explorer (JUICE), Dragonfly, and Europa Clipper missions are of highest priority, as JUICE and Europa Clipper have launched and Dragonfly is scheduled to arrive at its destination by the early 2030s.

\paragraph{Joint proposal opportunities with external facilities should be retained.}\label{par:joint}%
The opportunity to propose for joint observations with external
facilities reduces ``double jeopardy'' aspects related to having to
submit two or more proposals to different institutions to pursue a
single science goal, and it likely has the effects of encouraging
observations with potentially higher impact and lowering the
``potential barrier'' for individuals who do not have a long history
of observing with the \hbox{VLA}.

\subsection{Assessments of Technical Options}\label{sec:find.assess}

We now provide brief rationales or motivations for the specific
assessments in Table~\ref{tab:tmatrix}.
In addition to these science-focused considerations, this section closes with a discussion of potential cost savings associated these technical options that reflect feedback received by the ITAT.

\subsubsection{Reduced Receiver Suite}

Consistent with the more general finding of the uniqueness of the VLA
and VLBA's spectral dynamic range, a reduction of the VLA's receiver suite
accomplished by eliminating the VLA's
low- or high-frequency capabilities would not be acceptable during the VLA+VLBA to ngVLA transition. Instead, reducing the number of frequencies that could be observed, while
maintaining approximately the full spectral dynamic range of the
\hbox{VLA}, would ensure that the VLA continues to make
``very significant'' contributions to addressing the \textit{Pathways
  to Discovery} science priority questions, while also contributing in a supporting role to many others. Indeed:
\begin{itemize}
\item%
  A key aspect of understanding stellar physics, particularly in its relation to any orbiting planets, is magnetically-driven activity.  Moreover, such observations are ones for which radio wavelengths provide critical information.  At the Sun, tracing magnetic field structures is conducted by tracking emissions as a function of frequency, which requires observations from at least K~band to S~band.

\item%
  Many explosive phenomena produce incoherent emission (and sometimes are termed ``slow transients,'' e.g., novae, Type~II supernovae, gamma-ray burst afterglows), which is distinguished by a spectrum with characteristic frequencies that evolve to lower values with time, and specifically with a self-absorption frequency that cascades to lower values as the emitting region becomes increasingly optically thin.  Tracking the self-absorption frequency evolution, and more generally constraining the position of the break frquencies via panchromatic observations including the radio band, provides unparalleled (or nearly so) measures of the explosion parameters such as energy of the ejecta and interstellar medium density.

\item%
  As material moves along a jet, it can and often does evolve in frequency.  Tracking this frequency evolution, including potential rebrightenings, as jets impact ambient material, provides constraints on both the jet itself and its environment.
\end{itemize}

\subsubsection{Restricted Array Configurations}\label{sec:find.assess.config}
The VLA's current operations include cycling through four
configurations, from a compact D configuration that is intended for
observations requiring high surface brightness sensitivities to an extended A
configuration that is intended for observations requiring high
angular resolution.  In general, restricting the VLA to only compact
configurations was assessed to be unacceptable during the VLA/VLBA to
ngVLA transition.
Particularly for the science priority questions for which the VLA
provides ``very significant'' contributions, the sources being
observed tend to be compact, and potentially faint, so high angular
resolution is required both to avoid ``blending'' with other sources
and to mitigate confusion due to faint, unresolved sources. Adding to these general considerations are the following important facts:
\begin{itemize}
\item%
  For nearby stellar
  systems observed at high angular resolution, which requires the
  extended configurations, it is often possible to resolve
  binaries.
  The consequence of not having compact configurations
  would be a reduced capability to observe the Sun, though some of
  those observations might be able to be provided by other
  observatories, such as the Extended Owens Valley Solar Array
  (EOVSA).

\item%
  A typical metric is
  that approximately arcsecond resolution is required to associate an
  extragalactic transient with its host while distinguishing host
  light from transient emission.
  Arcsecond resolution is also required to associate a radio
  counterpart with a transient observed at other wavelengths.
  For some observations of transients powered by explosive phenomena, a configuration that would combine some of the antenna locations from the B configuration with some of the antenna locations from the D configuration (a ``B+D'' configuration) could be acceptable, as it would provide for arcsecond- or sub-arcsecond--resolution at the higher frequencies.

\item%
  Jets likely are launched on scales of order of~10 gravitational
  radii or less.
  Jets evolve on scales of astronomical units to parsecs, and their effects on the ambient media or host galaxies can occur on sub-parsec to parsec scales.
  Consequently, sub-arcsecond angular resolutions are required.
\end{itemize}

The TAG discussed various ``mixed'' configurations, e.g., an ``A+C'' configuration that would combine some of the antenna locations from the A~configuration with some of the antenna locations from the C~configuration, with the intent of providing arcsecond- or sub-arcsecond--resolution at the higher frequencies while preserving some measure of surface brightness sensitivity.
The TAG also was presented with the concept of an ``F'' configuration \cite{Fconfiguration}. Further analysis of this configuration requested to the ITAT by the TAG resulted in the ngVLA Memorandum~126 \citep{ngvla126}. This Memorandum highlights that it is difficult to identify a single configuration that preserves both the surface brightness sensitivity and the high angular resolution capabilities of the current \hbox{VLA}.
Further analysis by the ITAT also identified that the VLA operations are structured around being able to conduct regular maintenance on the antenna by rotating them through the ``antenna barn'' during configuration changes. Hence, with a fixed array configuration there would be an increased risk that required maintenance would be more difficult (or even could not be performed in the field), resulting both in reduced imaging performance (``poorer $u$-$v$ coverage'') and reduced sensitivity (\S\ref{sec:find.assess.number}). 
The consequences would be poorer performance for all science cases.

\subsubsection{Restricted Correlator Setups}\label{sec:find.assess.correlator}
The ITAT identified the elimination of the 3-bit sampler system from the VLA's correlator as the only possible benefit (in terms of reducing the required support effort) from a reduction in correlator setups.
The TAG did not consider this option at length.

\subsubsection{Reduced Number of Antennas}\label{sec:find.assess.number}
% reduced number of antennas
Current VLA operations require that no more than three antennas can be
out of service. 
For both the science priority questions for which the VLA can provide
``very significant'' contributions and across the larger set of ngVLA
science cases, reducing the number of VLA antennas in use was assessed
to have at least moderate and more likely significant effects.
Observations at radio wavelengths of polarized emission traditionally
have been one of the few means of tracing magnetic fields, which is of
cross-cutting relevance to the physics of solar and stellar phenomena,
explosive (transient) phenomena, and jets.
However, polarized
emission also typically is only a (small) fraction of the total
intensity, and a reduction in VLA sensitivity would be of
greater effect.
Moreover, observations of solar and stellar phenomena, explosive
(transient) phenomena, and jets often involve an element of the time
domain, tracking or monitoring changes in sources.
The reduction in sensitivity resulting from having fewer antennas
cannot be compensated by having longer-duration observations.
Indeed, the duration of some current observations are already
problematic or prohibitive.  For example, a single recent observation of the
late-time ejecta of GW170817 required 30~hrs.  If only 20 antennas were
available, 50~hr would have been required, and, if only 15 antennas
were available, more than 75~hr would have been required.
Finally, allowing for a reduced number of antennas affects not only
the sensitivity, but also potentially the configuration.  Depending
upon which antennas are out of service, or allowed to be out of
service temporarily, the resulting effective configuration might not
provide the angular resolution required for the various science cases.

\subsubsection{Reduced Observing Time}
% reduced observing time
Reducing the available observing time was assessed to have effects
ranging from moderate to significant.
\begin{itemize}
\item%
  While reduced observing likely would have a moderate effect on
  stellar observations, there would be a significant effect on solar
  observations if, for instance, no daytime observations were
  possible.  The effect might be mitigated somewhat because
  other facilities would be able to acquire at least some of the
  relevant observations.
\item%
  In general, explosive phenomena occur at random times.  A reduction
  in observing time introduces the significant risk that high-profile
  observations could not be acquired, particularly if the reduced
  observing time meant that simultaneous or contemporaneous
  observations with other facilities as part of a larger
  multi-wavelength campaign could not be conducted.

\item%
  Across the range of ngVLA Science Cases related to observing jets,
  the consequences of reduced observing time was judged to be of
  moderate impact.  The one potential exception is spectral timing of
  jets from stellar-mass compact objects; however, this science case
  also is covered by the assessment related to explosive phenomena.  
\end{itemize}

\subsubsection{Stop All VLA Observations}
% stopping VLA entirely
With regard to a complete cessation of VLA observations, there are
both scientific and foundational concerns.
From a scientific perspective, ceasing VLA observations for any
significant duration introduces the risk that it would not be able to
participate in multi-wavelength campaigns with other facilities in
pursuit of high-profile science.  Ceasing VLA observations for any
significant duration also may introduce the risk that current
agreements with other telescopes and facilities for joint proposals
could not be sustained.
From a foundational perspective, the absence of new VLA observations
likely would result in significant harm to early career researchers,
the cohort that will make the greatest use of the \hbox{ngVLA}.
The standard post-doctoral researcher appointments range from one to
three years.  A cessation of even six months would be a significant
fraction of a post-doctoral researcher's appointment.  Longer duration
cessations would have the potential to affect not only post-doctoral
researchers but graduate students as well.

While the TAG assessed that a cessation of VLA observations for less
than three months likely would be of low to moderate impact, both for
the science priority questions for which the VLA provides ``very
significant'' contributions and for the broader range of ngVLA Science
Cases, there is the risk that unforseen events would cause the
duration to increase well beyond three months.

\subsection{Cost Savings and Personnel Impacts}\label{sec:find.costs}

The charge to the TAG includes not only a science assessment of potential VLA+VLBA to ngVLA transition options (\S\ref{sec:charge}), but also ``their cost'' and ``technical/personnel impacts.''
Following the development of a draft initial set of recommendations, the TAG received from the NRAO's ITAT an assessment of the potential cost savings and implications for personnel of those recommendations.
The consensus within the TAG was that the projected cost savings, even if all of the potential options were adopted, was not significant, particularly with respect to the projected construction cost of the \hbox{ngVLA}. 
The ITAT's assessment did influence the priority ranking of the TAG's recommendations.

\clearpage
\section{Recommendations}\label{sec:recommend}

This Section presents the recommended approaches for the VLA+VLBA to ngVLA Transition. Recognizing that circumstances and scientific priorities may change between the time of this report and the start of the Transition itself, the TAG presents four recommended options, ordered by priority.

The TAG emphasizes that any reduction in capability, to either the VLA or the VLBA or both, will have scientific consequences, as demonstrated by the many references to work involving these telescopes that are cited in \textit{Pathways to Discovery} (and \textit{Origins, Worlds, and Life} and \textit{Exploring and Safeguarding Humanity's Home in Space}) and described in in~\S\ref{sec:science}.
The recommendations of the TAG are developed under the assumption that the Transition is both limited in duration and concludes with much greater capability, namely the ngVLA such as described in the NRAO and community white papers forming the basis of the support for the ngVLA in \textit{Pathways to Discovery}.

In that spirit, in order to maximize the scientific return during the VLA+VLBA to ngVLA transition, the TAG recommends that the \textit{transition instrument} that will bridge the operations of the VLA+VLBA to the ngVLA does not begin operations until ngVLA construction funds have been obligated and construction activities have begun.  The transition should end once ngVLA capabilities comparable to the current VLA/VLBA are operational.

\subsection{VLBA Transition}\label{sec:rec.VLBA}

The TAG recommends that the VLBA observational capabilities remain unchanged compared to current VLBA capabilities, including the full receiver suite, during the transition.
Moreover, any significant changes in the operational capabilities of the VLBA would have to be coordinated with the United States Naval Observatory (USNO), which provides 50\% of VLBA's operational funding (\S\ref{sec:other}).
The TAG did not consider this aspect of the VLBA operations in its deliberations.

\subsection{First VLA Transition Option}\label{sec:prefer}

Given the preeminent position of the VLA and the risks associated with a long duration, the TAG recommends that the VLA should remain at full capabilities for as long as possible during the construction of the \hbox{ngVLA}.
Specifically, with reference to Figure~\ref{fig:transition}, which is based on the current plan for the design, development, and construction of the \hbox{ngVLA} provided by \hbox{NRAO}, a reasonable transition plan is the following:  A three-year interval during the initial ngVLA construction during which the VLA capabilities remain consistent with current capabilities, followed by a two-year interval during which one or both of the transition options described below could be used concurrently with ngVLA Early Science.

The TAG acknowledges that this plan carries risks as well.  
As part of the NRAO/ITAT response to the TAG's initial set of recommendations, various subsystems within the VLA were identified as either at or approaching obsolescence.
Should one of those subsystems fail, and repair be infeasible or exorbitant, accelerating the Transition may be required.

\subsection{Second VLA Transition Option}\label{sec:alt2}

Should the first transition option be infeasible or insufficient, the recommended second option is to reduce the set of receivers at each antenna, provided that at least five of the current frequency bands are maintained at all antennas.
{The TAG judges that no fewer than five frequency bands are required in order to enable the VLA to have a sufficient spectral dynamic range} (\S\ref{sec:find}).
Further, at a minimum, these five frequency bands must span from ``low'' frequency to ``high'' frequency.  Here, ``low'' frequency is considered to be C~band (4~GHz--8~GHz) and lower frequencies while ``high'' frequency is considered to be K${}_u$~band (12~GHz--18~GHz) and higher frequencies.  
X~band (8~GHz--12~GHz) must be maintained at all times as it is required for reference pointing calibration for higher frequencies, and X~band is one of the most heavily-demanded bands on the \hbox{VLA}. Based on these considerations, the baseline set of receivers for the transition is \hbox{L-S-C-X-K}. We note that this suite of receivers does not include either the low-frequency (P-band) or Q-band receivers, which has the effect of reducing the extent to which the VLA’s frequency range is unique.

\subsection{Third VLA Transition Option}\label{sec:alt1}
Given that no single arrangement of antennas appears capable of retaining the VLA's combination of surface brightness sensitivity and angular resolution capabilities \citep{ngvla126}, and the potential for increased difficulties in maintaining the antennas (\S\ref{sec:find.assess.config}), a fixed configuration should be adopted only if necessary.

\clearpage

\section{Other Considerations}\label{sec:other}

In addition to the specific considerations of the scientific opportunities and technical options, the TAG discussed other topics related to the VLA+VLBA to ngVLA transition.

Experience from the transition from the VLA to the Expanded VLA (EVLA) and finally the Karl G.~Jansky VLA should be taken into account.  
Most notably, as commissioning of the ngVLA begins, involving the community in so-called ``shared risk observing'' ensures engagement and enables more rapid commissioning progress to be made.

While the TAG recommends that the Transition be of as limited duration as possible (\S\ref{sec:prefer}), there is the possibility that VLA observations would have to cease for some interval.
Even so, the VLA Archive remains a rich source of scientific results and a means by which early-career researchers can gain experience with radio astronomical interferometers.
The TAG did not discuss such possibilities at length, but a limited-duration ``archival data analysis'' program may be warranted to ensure continued engagement between the current NRAO user community and NRAO facilities, should there have to be a cessation of new VLA observations.

Finally, the USNO is a 50\% partner with the NSF in supporting the operations and maintenance (O\&M) of the \hbox{VLBA}.
The USNO requires daily, weekly, monthly, and other non-regular cadence use of VLBA in order to sustain its mission in support of producing the Celestial Reference Frame (CRF) and Earth Orientation Parameters (EOPs).
In addition to the O\&M funds, the USNO, working with its mission partners, has enabled the funding of significant upgrades to the VLBA including hydrogen masers, generators, new high-rate tracking receivers that are currently being deployed across all ten VLBA sites, and the VLBA New Digital Architecture (VNDA) upgrade, which will digitize the signal chain and replace the aging, outdated, irreplaceable Roach Digital Backends (RDBEs).
These modernization efforts are expensive investments that are made in order to ensure the resilience of the VLBA for use in USNO's CRF and EOP mission well into the 2030's timeframe.
Any transition from VLBA to ngVLA will require careful coordination with USNO in order to prevent any loss in the ability to meet USNO's mission requirements.

\clearpage
\section{Stakeholder Engagement}\label{sec:engage}

The TAG engaged in a number of activities in an effort to ensure broad awareness within the community and solicit input from the community.

\begin{itemize}
\item%
At the NRAO Town Hall during the $241^{\mathrm{st}}$~Meeting of
  the American Astronomical Society (2023 January), the ngVLA Project Scientist described the TAG and identified the Chairs.
  Multiple individuals subsequently discussed with the Chairs the
  TAG's charge and its importance.

\item%
A poster reporting the current status was presented at the ``New Eyes on the Universe'' conference sponsored by NRAO and the Square Kilometre Array Observatory (SKAO), and a ``poster flash'' talk also was given.

\item%
A Splinter Session entitled ``The Next-Generation VLA: Update and Community Forum'' was organized at the $242^{\mathrm{st}}$~Meeting of the American Astronomical Society (2023 June) in order to present an overview of the current status and seek community input.

\item%
The NRAO Users Committee has been briefed by NRAO on the progress of the \hbox{TAG}.  Among comments in published Users Committee reports are an interest in ensuring that simulations are conducted prior to any decision to eliminate VLA configuration changes, echoing a TAG recommendation.

\item%
The NRAO Newsletter (Volume~16, Issue~6, 2023 June~23) contained a summary of the Splinter Session at the $242^{\mathrm{st}}$~AAS Meeting, including a link to the slides presented and an invitation to submit feedback via an online form.

\item%
A poster reporting the current status was presented at the ``Follow the Monarchs: A Journey to Explore the Cosmos at (Sub)milliarcsecond Scales with the ngVLA'' conference, and a ``poster flash'' talk also was given.

\item%
A Splinter Session entitled ``The Next-Generation VLA: Update and Community Forum'' was organized at the $245^{\mathrm{th}}$~Meeting of the American Astronomical Society (2025 January) in order to present an overview of the current status and seek community input.
The TAG has posted its initial report on arXiv (astro-ph) for a 90-day comment period  for community members to submit comments.  These comments have been addressed in this final version of the report.
\end{itemize}

\clearpage

\bibliographystyle{unsrt}
%\bibliography{references.bib}

\begin{thebibliography}{100}

\bibitem{astro2020}
{National Academies of Sciences, Engineering, and Medicine}.
\newblock {\em {\textit{Pathways to Discovery in Astronomy and Astrophysics for the 2020s}}}.
\newblock {The National Academies Press}, {Washington, DC}, 2021.

\bibitem{origin2023}
{National Academies of Sciences, Engineering, and Medicine}.
\newblock {\em {\textit{Origins, Worlds, and Life: A Decadal Strategy for Planetary Science and Astrobiology 2023--2032}}}.
\newblock {The National Academies Press}, {Washington, DC}, 2023.

\bibitem{helio}
{National Academies of Sciences, Engineering, and Medicine}.
\newblock {\em {\textit{The Next Decade of Discovery in Solar and Space Physics: Exploring and Safeguarding Humanity's Home in Space}}}.
\newblock {The National Academies Press}, {Washington, DC}, 2024.

\bibitem{ngVLAScienceBook}
{Eric Murphy and the ngVLA Science Advisory Council}, editor.
\newblock {\em {Science with a Next Generation Very Large Array}}, volume 517 of {\em Astronomical Society of the Pacific Conference Series}.
\newblock Astronomical Society of the Pacific, 2018.

\bibitem{GW170817press}
Radio eyes unlocking secrets of neutron-star collision.
\newblock \url{https://public.nrao.edu/news/radio-eyes-unlocking-secrets/}, 2017.

\bibitem{Abbott2017a}
{LIGO Scientific Collaboration} and {Virgo Collaboration}.
\newblock {GW170817: Observation of Gravitational Waves from a Binary Neutron Star Inspiral}.
\newblock {\em PRL}, 119(16):161101, October 2017.

\bibitem{Abbott2017b}
{LIGO Scientific Collaboration}, {Virgo Collaboration}, et~al.
\newblock {Multi-messenger Observations of a Binary Neutron Star Merger}.
\newblock {\em ApJL}, 848(2):L12, October 2017.

\bibitem{Aartsen2018}
{IceCube Collaboration}.
\newblock {Multimessenger observations of a flaring blazar coincident with high-energy neutrino IceCube-170922A}.
\newblock {\em Science}, 361(6398):eaat1378, July 2018.

\bibitem{Alexander2017}
K.~D. {Alexander}, E.~{Berger}, W.~{Fong}, P.~K.~G. {Williams}, C.~{Guidorzi}, R.~{Margutti}, B.~D. {Metzger}, J.~{Annis}, P.~K. {Blanchard}, D.~{Brout}, D.~A. {Brown}, H.~Y. {Chen}, R.~{Chornock}, P.~S. {Cowperthwaite}, M.~{Drout}, T.~{Eftekhari}, J.~{Frieman}, D.~E. {Holz}, M.~{Nicholl}, A.~{Rest}, M.~{Sako}, M.~{Soares-Santos}, and V.~A. {Villar}.
\newblock {The Electromagnetic Counterpart of the Binary Neutron Star Merger LIGO/Virgo GW170817. VI.~Radio Constraints on a Relativistic Jet and Predictions for Late-time Emission from the Kilonova Ejecta}.
\newblock {\em ApJL}, 848(2):L21, October 2017.

\bibitem{Hallinan2017}
G.~{Hallinan}, A.~{Corsi}, K.~P. {Mooley}, K.~{Hotokezaka}, E.~{Nakar}, M.~M. {Kasliwal}, D.~L. {Kaplan}, D.~A. {Frail}, S.~T. {Myers}, T.~{Murphy}, K.~{De}, D.~{Dobie}, J.~R. {Allison}, K.~W. {Bannister}, V.~{Bhalerao}, P.~{Chandra}, T.~E. {Clarke}, S.~{Giacintucci}, A.~Y.~Q. {Ho}, A.~{Horesh}, N.~E. {Kassim}, S.~R. {Kulkarni}, E.~{Lenc}, F.~J. {Lockman}, C.~{Lynch}, D.~{Nichols}, S.~{Nissanke}, N.~{Palliyaguru}, W.~M. {Peters}, T.~{Piran}, J.~{Rana}, E.~M. {Sadler}, and L.~P. {Singer}.
\newblock {A radio counterpart to a neutron star merger}.
\newblock {\em Science}, 358(6370):1579--1583, December 2017.

\bibitem{Lazzati2018}
Davide {Lazzati}, Rosalba {Perna}, Brian~J. {Morsony}, Diego {Lopez-Camara}, Matteo {Cantiello}, Riccardo {Ciolfi}, Bruno {Giacomazzo}, and Jared~C. {Workman}.
\newblock {Late Time Afterglow Observations Reveal a Collimated Relativistic Jet in the Ejecta of the Binary Neutron Star Merger GW170817}.
\newblock {\em PRL}, 120(24):241103, June 2018.

\bibitem{Mooley2018a}
K.~P. {Mooley}, E.~{Nakar}, K.~{Hotokezaka}, G.~{Hallinan}, A.~{Corsi}, D.~A. {Frail}, A.~{Horesh}, T.~{Murphy}, E.~{Lenc}, D.~L. {Kaplan}, K.~{de}, D.~{Dobie}, P.~{Chandra}, A.~{Deller}, O.~{Gottlieb}, M.~M. {Kasliwal}, S.~R. {Kulkarni}, S.~T. {Myers}, S.~{Nissanke}, T.~{Piran}, C.~{Lynch}, V.~{Bhalerao}, S.~{Bourke}, K.~W. {Bannister}, and L.~P. {Singer}.
\newblock {A mildly relativistic wide-angle outflow in the neutron-star merger event GW170817}.
\newblock {\em Nature}, 554(7691):207--210, February 2018.

\bibitem{Hovatta2021}
T.~{Hovatta}, E.~{Lindfors}, S.~{Kiehlmann}, W.~{Max-Moerbeck}, M.~{Hodges}, I.~{Liodakis}, A.~{L{\"a}hteem{\"a}ki}, T.~J. {Pearson}, A.~C.~S. {Readhead}, R.~A. {Reeves}, S.~{Suutarinen}, J.~{Tammi}, and M.~{Tornikoski}.
\newblock {Association of IceCube neutrinos with radio sources observed at Owens Valley and Mets{\"a}hovi Radio Observatories}.
\newblock {\em A\&A}, 650:A83, June 2021.

\bibitem{Balasubramanian2022}
Arvind {Balasubramanian}, Alessandra {Corsi}, Kunal~P. {Mooley}, Kenta {Hotokezaka}, David~L. {Kaplan}, Dale~A. {Frail}, Gregg {Hallinan}, Davide {Lazzati}, and Eric~J. {Murphy}.
\newblock {GW170817 4.5~Yr After Merger: Dynamical Ejecta Afterglow Constraints}.
\newblock {\em ApJ}, 938(1):12, October 2022.

\bibitem{2020ApJ...894..101P}
Alexander {Plavin}, Yuri~Y. {Kovalev}, Yuri~A. {Kovalev}, and Sergey {Troitsky}.
\newblock {Observational Evidence for the Origin of High-energy Neutrinos in Parsec-scale Nuclei of Radio-bright Active Galaxies}.
\newblock {\em ApJ}, 894(2):101, 2020.

\bibitem{2021ApJ...908..157P}
A.~V. {Plavin}, Y.~Y. {Kovalev}, Yu.~A. {Kovalev}, and S.~V. {Troitsky}.
\newblock {Directional Association of TeV to PeV Astrophysical Neutrinos with Radio Blazars}.
\newblock {\em ApJ}, 908(2):157, February 2021.

\bibitem{2024MNRAS.527L..26S}
Alisa {Suray} and Sergey {Troitsky}.
\newblock {Neutrino flares of radio blazars observed from TeV to PeV}.
\newblock {\em MNRAS}, 527(1):L26--L31, January 2024.

\bibitem{GW170817press2}
Radio observations confirm superfast jet of material from neutron star merger.
\newblock \url{https://public.nrao.edu/news/superfast-jet-neutron-star-merger/}, 2017.

\bibitem{Mooley2018b}
K.~P. {Mooley}, A.~T. {Deller}, O.~{Gottlieb}, E.~{Nakar}, G.~{Hallinan}, S.~{Bourke}, D.~A. {Frail}, A.~{Horesh}, A.~{Corsi}, and K.~{Hotokezaka}.
\newblock {Superluminal motion of a relativistic jet in the neutron-star merger GW170817}.
\newblock {\em Nature}, 561(7723):355--359, September 2018.

\bibitem{Ghirlanda2019}
G.~{Ghirlanda}, O.~S. {Salafia}, Z.~{Paragi}, M.~{Giroletti}, J.~{Yang}, B.~{Marcote}, J.~{Blanchard}, I.~{Agudo}, T.~{An}, M.~G. {Bernardini}, R.~{Beswick}, M.~{Branchesi}, S.~{Campana}, C.~{Casadio}, E.~{Chassande-Mottin}, M.~{Colpi}, S.~{Covino}, P.~{D'Avanzo}, V.~{D'Elia}, S.~{Frey}, M.~{Gawronski}, G.~{Ghisellini}, L.~I. {Gurvits}, P.~G. {Jonker}, H.~J. {van Langevelde}, A.~{Melandri}, J.~{Moldon}, L.~{Nava}, A.~{Perego}, M.~A. {Perez-Torres}, C.~{Reynolds}, R.~{Salvaterra}, G.~{Tagliaferri}, T.~{Venturi}, S.~D. {Vergani}, and M.~{Zhang}.
\newblock {Compact radio emission indicates a structured jet was produced by a binary neutron star merger}.
\newblock {\em Science}, 363(6430):968--971, March 2019.

\bibitem{LHAASO2023}
{LHAASO Collaboration}.
\newblock {A tera-electron volt afterglow from a narrow jet in an extremely bright gamma-ray burst.}
\newblock {\em Science}, 380(6652):1390--1396, June 2023.

\bibitem{2015ApJ...808..189W}
P.~K.~G. {Williams} and E.~{Berger}.
\newblock {The Rotation Period and Magnetic Field of the T Dwarf 2MASSI J1047539+212423 Measured from Periodic Radio Bursts}.
\newblock {\em ApJ}, 808(2):189, August 2015.

\bibitem{2017A&A...599A.127F}
Bibiana {Fichtinger}, Manuel {G{\"u}del}, Robert~L. {Mutel}, Gregg {Hallinan}, Eric {Gaidos}, Stephen~L. {Skinner}, Christene {Lynch}, and Kenneth~G. {Gayley}.
\newblock {Radio emission and mass loss rate limits of four young solar-type stars}.
\newblock {\em A\&A}, 599:A127, March 2017.

\bibitem{2017A&A...602A..57A}
R.~{Azulay}, J.~C. {Guirado}, J.~M. {Marcaide}, I.~{Mart{\'\i}-Vidal}, E.~{Ros}, E.~{Tognelli}, F.~{Hormuth}, and J.~L. {Ortiz}.
\newblock {Young, active radio stars in the AB Doradus moving group}.
\newblock {\em A\&A}, 602:A57, June 2017.

\bibitem{2018A&A...612A..48W}
K.~T. {Wong}, K.~M. {Menten}, T.~{Kami{\'n}ski}, F.~{Wyrowski}, J.~H. {Lacy}, and T.~K. {Greathouse}.
\newblock {Circumstellar ammonia in oxygen-rich evolved stars}.
\newblock {\em A\&A}, 612:A48, April 2018.

\bibitem{2023AJ....165...92M}
L.~D. {Matthews}, N.~R. {Evans}, and M.~P. {Rupen}.
\newblock {First Detection of Radio Emission Associated with a Classical Cepheid}.
\newblock {\em Astrophys.J.}, 165(3):92, March 2023.

\bibitem{2021ApJ...911....4L}
Yingjie {Luo}, Bin {Chen}, Sijie {Yu}, T.~S. {Bastian}, and S{\"a}m {Krucker}.
\newblock {Radio Spectral Imaging of an M8.4 Eruptive Solar Flare: Possible Evidence of a Termination Shock}.
\newblock {\em ApJ}, 911(1):4, April 2021.

\bibitem{2022ApJ...938...12B}
Arvind {Balasubramanian}, Alessandra {Corsi}, Kunal~P. {Mooley}, Kenta {Hotokezaka}, David~L. {Kaplan}, Dale~A. {Frail}, Gregg {Hallinan}, Davide {Lazzati}, and Eric~J. {Murphy}.
\newblock {GW170817 4.5 Yr After Merger: Dynamical Ejecta Afterglow Constraints}.
\newblock {\em ApJ}, 938(1):12, October 2022.

\bibitem{2018ApJ...861L..10C}
Alessandra {Corsi}, Gregg~W. {Hallinan}, Davide {Lazzati}, Kunal~P. {Mooley}, Eric~J. {Murphy}, Dale~A. {Frail}, Dario {Carbone}, David~L. {Kaplan}, Tara {Murphy}, Shrinivas~R. {Kulkarni}, and Kenta {Hotokezaka}.
\newblock {An Upper Limit on the Linear Polarization Fraction of the GW170817 Radio Continuum}.
\newblock {\em ApJL}, 861(1):L10, July 2018.

\bibitem{mooley2018}
K.~P. {Mooley}, A.~T. {Deller}, O.~{Gottlieb}, E.~{Nakar}, G.~{Hallinan}, S.~{Bourke}, D.~A. {Frail}, A.~{Horesh}, A.~{Corsi}, and K.~{Hotokezaka}.
\newblock {Superluminal motion of a relativistic jet in the neutron-star merger GW170817}.
\newblock {\em Nature}, 561(7723):355--359, September 2018.

\bibitem{Ioka2018}
Kunihito {Ioka} and Takashi {Nakamura}.
\newblock {Can an off-axis gamma-ray burst jet in GW170817 explain all the electromagnetic counterparts?}
\newblock {\em Progress of Theoretical and Experimental Physics}, 2018(4):043E02, April 2018.

\bibitem{2018ApJ...861...12L}
M.~T. {Lam}, M.~A. {McLaughlin}, J.~M. {Cordes}, S.~{Chatterjee}, and T.~J.~W. {Lazio}.
\newblock {Optimal Frequency Ranges for Submicrosecond Precision Pulsar Timing}.
\newblock {\em ApJ}, 861(1):12, July 2018.

\bibitem{2013MNRAS.433.3133G}
I.~{Gonidakis}, P.~J. {Diamond}, and A.~J. {Kemball}.
\newblock {A long-term VLBA monitoring campaign of the v = 1, J = 1 {\textrightarrow}0 SiO masers towards TX Cam - I. Morphology and shock waves}.
\newblock {\em MNRAS}, 433(4):3133--3151, August 2013.

\bibitem{2018ApJ...853...42S}
J.~B. {Su}, Z.-Q. {Shen}, X.~{Chen}, and D.~R. {Jiang}.
\newblock {Dynamics of SiO Masers around VX~Sgr}.
\newblock {\em ApJ}, 853(1):42, January 2018.

\bibitem{2012A&A...544A..42Z}
B.~{Zhang}, M.~J. {Reid}, K.~M. {Menten}, X.~W. {Zheng}, and A.~{Brunthaler}.
\newblock {The distance and size of the red hypergiant NML Cygni from VLBA and VLA astrometry}.
\newblock {\em A\&A}, 544:A42, August 2012.

\bibitem{SupergiantPress}
Supergiant atmosphere of antares revealed by radio telescopes.
\newblock \url{https://public.nrao.edu/news/supergiant-atmosphere-of-antares-revealed-by-radio-telescopes/}, 2020.

\bibitem{2020A&A...638A..65O}
E.~{O'Gorman}, G.~M. {Harper}, K.~{Ohnaka}, A.~{Feeney-Johansson}, K.~{Wilkeneit-Braun}, A.~{Brown}, E.~F. {Guinan}, J.~{Lim}, A.~M.~S. {Richards}, N.~{Ryde}, and W.~H.~T. {Vlemmings}.
\newblock {ALMA and VLA reveal the lukewarm chromospheres of the nearby red supergiants Antares and Betelgeuse}.
\newblock {\em A\&A}, 638:A65, June 2020.

\bibitem{BrownDwarfPress}
Astronomers measure wind speed on a brown dwarf.
\newblock \url{https://public.nrao.edu/news/brown-dwarf-wind-speed/}, 2020.

\bibitem{2020Sci...368..169A}
Katelyn.~N. {Allers}, Johanna~M. {Vos}, Beth~A. {Biller}, and Peter. K.~G. {Williams}.
\newblock {A measurement of the wind speed on a brown dwarf}.
\newblock {\em Science}, 368(6487):169--172, April 2020.

\bibitem{VLBAPlanetPress}
Vlba finds planet orbiting small, cool star.
\newblock \url{https://public.nrao.edu/news/vlba-finds-planet/}, 2020.

\bibitem{VLBAPlanetPress2}
Vlba produces first full 3-d view of binary star-planet system.
\newblock \url{https://public.nrao.edu/news/binary-star-planet-system/}, 2022.

\bibitem{2016ApJ...827...22F}
Jan {Forbrich}, Trent~J. {Dupuy}, Mark~J. {Reid}, Edo {Berger}, Aaron {Rizzuto}, Andrew~W. {Mann}, Michael~C. {Liu}, Kimberly {Aller}, and Adam~L. {Kraus}.
\newblock {High-precision Radio and Infrared Astrometry of LSPM J1314+1320AB. I. Parallax, Proper Motions, and Limits on Planets}.
\newblock {\em ApJ}, 827(1):22, August 2016.

\bibitem{2016ApJ...827...23D}
Trent~J. {Dupuy}, Jan {Forbrich}, Aaron {Rizzuto}, Andrew~W. {Mann}, Kimberly {Aller}, Michael~C. {Liu}, Adam~L. {Kraus}, and Edo {Berger}.
\newblock {High-precision Radio and Infrared Astrometry of LSPM J1314+1320AB. II. Testing Pre-main-sequence Models at the Lithium Depletion Boundary with Dynamical Masses}.
\newblock {\em ApJ}, 827(1):23, August 2016.

\bibitem{Curiel_2020}
Salvador {Curiel}, Gisela~N. {Ortiz-Le{\'o}n}, Amy~J. {Mioduszewski}, and Rosa~M. {Torres}.
\newblock {An Astrometric Planetary Companion Candidate to the M9 Dwarf TVLM~513-46546}.
\newblock {\em AJ}, 160(3):97, September 2020.

\bibitem{2024ApJ...967..112C}
Salvador {Curiel}, Gisela~N. {Ortiz-Le{\'o}n}, Amy~J. {Mioduszewski}, and Anthony~B. {Arenas-Martinez}.
\newblock {Precise Mass, Orbital Motion, and Stellar Properties of the M-dwarf Binary LP 349‑25AB}.
\newblock {\em ApJ}, 967(2):112, June 2024.

\bibitem{2020ApJ...892...82R}
Luis~F. {Rodr{\'\i}guez}, Sergio~A. {Dzib}, Luis {Zapata}, Susana {Lizano}, Laurent {Loinard}, Karl~M. {Menten}, and Laura {G{\'o}mez}.
\newblock {Proper Motions of the Radio Source Orion~\hbox{MR}, Formerly Known as Orion~n, and New Sources with Large Proper Motions in Orion BN/KL}.
\newblock {\em ApJ}, 892(2):82, April 2020.

\bibitem{2021ARA&A..59...21G}
Suvi {Gezari}.
\newblock {Tidal Disruption Events}.
\newblock {\em ARA\&A}, 59:21--58, September 2021.

\bibitem{SupernovaPress}
Stellar collision triggers supernova explosion.
\newblock \url{https://public.nrao.edu/news/stellar-collision-triggers-supernova/}, 2021.

\bibitem{2022ApJ...924...16M}
K.~P. {Mooley}, B.~{Margalit}, C.~J. {Law}, D.~A. {Perley}, A.~T. {Deller}, T.~J.~W. {Lazio}, M.~F. {Bietenholz}, T.~{Shimwell}, H.~T. {Intema}, B.~M. {Gaensler}, B.~D. {Metzger}, D.~Z. {Dong}, G.~{Hallinan}, E.~O. {Ofek}, and L.~{Sironi}.
\newblock {Late-time Evolution and Modeling of the Off-axis Gamma-Ray Burst Candidate FIRST J141918.9+394036}.
\newblock {\em ApJ}, 924(1):16, January 2022.

\bibitem{FRBPress}
Strange radio burst raises new questions.
\newblock \url{https://public.nrao.edu/news/strange-radio-burst-raises-new-questions/}, 2022.

\bibitem{2019ApJ...886...24L}
C.~J. {Law}, C.~M.~B. {Omand}, K.~{Kashiyama}, K.~{Murase}, G.~C. {Bower}, K.~{Aggarwal}, S.~{Burke-Spolaor}, B.~J. {Butler}, P.~{Demorest}, T.~J.~W. {Lazio}, J.~{Linford}, S.~P. {Tendulkar}, and M.~P. {Rupen}.
\newblock {A Search for Late-time Radio Emission and Fast Radio Bursts from Superluminous Supernovae}.
\newblock {\em ApJ}, 886(1):24, November 2019.

\bibitem{2020ApJ...899..161L}
Casey~J. {Law}, Bryan~J. {Butler}, J.~Xavier {Prochaska}, Barak {Zackay}, Sarah {Burke-Spolaor}, Alexandra {Mannings}, Nicolas {Tejos}, Alexander {Josephy}, Bridget {Andersen}, Pragya {Chawla}, Kasper~E. {Heintz}, Kshitij {Aggarwal}, Geoffrey~C. {Bower}, Paul~B. {Demorest}, Charles~D. {Kilpatrick}, T.~Joseph~W. {Lazio}, Justin {Linford}, Ryan {Mckinven}, Shriharsh {Tendulkar}, and Sunil {Simha}.
\newblock {A Distant Fast Radio Burst Associated with Its Host Galaxy by the Very Large Array}.
\newblock {\em ApJ}, 899(2):161, August 2020.

\bibitem{2022Natur.606..873N}
C.~H. {Niu}, K.~{Aggarwal}, D.~{Li}, X.~{Zhang}, S.~{Chatterjee}, C.~W. {Tsai}, W.~{Yu}, C.~J. {Law}, S.~{Burke-Spolaor}, J.~M. {Cordes}, Y.~K. {Zhang}, S.~K. {Ocker}, J.~M. {Yao}, P.~{Wang}, Y.~{Feng}, Y.~{Niino}, C.~{Bochenek}, M.~{Cruces}, L.~{Connor}, J.~A. {Jiang}, S.~{Dai}, R.~{Luo}, G.~D. {Li}, C.~C. {Miao}, J.~R. {Niu}, R.~{Anna-Thomas}, J.~{Sydnor}, D.~{Stern}, W.~Y. {Wang}, M.~{Yuan}, Y.~L. {Yue}, D.~J. {Zhou}, Z.~{Yan}, W.~W. {Zhu}, and B.~{Zhang}.
\newblock {A repeating fast radio burst associated with a persistent radio source}.
\newblock {\em Nature}, 606(7916):873--877, June 2022.

\bibitem{CBlastPress}
Astronomers study mysterious new type of cosmic blast.
\newblock \url{https://public.nrao.edu/news/mysterious-cosmic-blast/}, 2019.

\bibitem{CExplosionPress}
Astronomers discover new class of cosmic explosions.
\newblock \url{https://public.nrao.edu/news/new-class-cosmic-explosions/}, 2020.

\bibitem{2020ApJ...895L..23C}
D.~L. {Coppejans}, R.~{Margutti}, G.~{Terreran}, A.~J. {Nayana}, E.~R. {Coughlin}, T.~{Laskar}, K.~D. {Alexander}, M.~{Bietenholz}, D.~{Caprioli}, P.~{Chandra}, M.~R. {Drout}, D.~{Frederiks}, C.~{Frohmaier}, K.~H. {Hurley}, C.~S. {Kochanek}, M.~{MacLeod}, A.~{Meisner}, P.~E. {Nugent}, A.~{Ridnaia}, D.~J. {Sand}, D.~{Svinkin}, C.~{Ward}, S.~{Yang}, A.~{Baldeschi}, I.~V. {Chilingarian}, Y.~{Dong}, C.~{Esquivia}, W.~{Fong}, C.~{Guidorzi}, P.~{Lundqvist}, D.~{Milisavljevic}, K.~{Paterson}, D.~E. {Reichart}, B.~{Shappee}, M.~C. {Stroh}, S.~{Valenti}, B.~A. {Zauderer}, and B.~{Zhang}.
\newblock {A Mildly Relativistic Outflow from the Energetic, Fast-rising Blue Optical Transient CSS161010 in a Dwarf Galaxy}.
\newblock {\em ApJL}, 895(1):L23, 2020.

\bibitem{2020ApJ...895...49H}
Anna Y.~Q. {Ho}, Daniel~A. {Perley}, S.~R. {Kulkarni}, Dillon Z.~J. {Dong}, Kishalay {De}, Poonam {Chandra}, Igor {Andreoni}, Eric~C. {Bellm}, Kevin~B. {Burdge}, Michael {Coughlin}, Richard {Dekany}, Michael {Feeney}, Dmitry~D. {Frederiks}, Christoffer {Fremling}, V.~Zach {Golkhou}, Matthew~J. {Graham}, David {Hale}, George {Helou}, Assaf {Horesh}, Mansi~M. {Kasliwal}, Russ~R. {Laher}, Frank~J. {Masci}, A.~A. {Miller}, Michael {Porter}, Anna {Ridnaia}, Ben {Rusholme}, David~L. {Shupe}, Maayane~T. {Soumagnac}, and Dmitry~S. {Svinkin}.
\newblock {The Koala: A Fast Blue Optical Transient with Luminous Radio Emission from a Starburst Dwarf Galaxy at z = 0.27}.
\newblock {\em ApJ}, 895(1):49, 2020.

\bibitem{VLASSPress}
Vla sky survey reveals first 'orphan' gamma ray burs.
\newblock \url{https://public.nrao.edu/news/orphan-gamma-ray-burst/}, 2018.

\bibitem{2018ApJ...866L..22L}
C.~J. {Law}, B.~M. {Gaensler}, B.~D. {Metzger}, E.~O. {Ofek}, and L.~{Sironi}.
\newblock {Discovery of the Luminous, Decades-long, Extragalactic Radio Transient FIRST J141918.9+394036}.
\newblock {\em ApJL}, 866(2):L22, October 2018.

\bibitem{NeutrinoPress}
Vla gives tantalizing clues about source of energetic cosmic neutrino.
\newblock \url{https://public.nrao.edu/news/vla-cosmic-neutrino/}, 2018.

\bibitem{2018Sci...361.1378I}
{IceCube Collaboration}, M.~G. {Aartsen}, M.~{Ackermann}, J.~{Adams}, J.~A. {Aguilar}, M.~{Ahlers}, M.~{Ahrens}, I.~{Al Samarai}, D.~{Altmann}, K.~{Andeen}, and et~al.
\newblock {Multimessenger observations of a flaring blazar coincident with high-energy neutrino IceCube-170922A}.
\newblock {\em Science}, 361(6398):eaat1378, July 2018.

\bibitem{2019Natur.569..374M}
James C.~A. {Miller-Jones}, Alexandra~J. {Tetarenko}, Gregory~R. {Sivakoff}, Matthew~J. {Middleton}, Diego {Altamirano}, Gemma~E. {Anderson}, Tomaso~M. {Belloni}, Rob~P. {Fender}, Peter~G. {Jonker}, Elmar~G. {K{\"o}rding}, Hans~A. {Krimm}, Dipankar {Maitra}, Sera {Markoff}, Simone {Migliari}, Kunal~P. {Mooley}, Michael~P. {Rupen}, David~M. {Russell}, Thomas~D. {Russell}, Craig~L. {Sarazin}, Roberto {Soria}, and Valeriu {Tudose}.
\newblock {A rapidly changing jet orientation in the stellar-mass black-hole system V404 Cygni}.
\newblock {\em Nature}, 569(7756):374--377, April 2019.

\bibitem{2021ApJ...923L...5P}
Alice {Pasetto}, Carlos {Carrasco-Gonz{\'a}lez}, Jos{\'e}~L. {G{\'o}mez}, Jos{\'e}-Maria {Mart{\'\i}}, Manel {Perucho}, Shane~P. {O'Sullivan}, Craig {Anderson}, Daniel~Jacobo {D{\'\i}az-Gonz{\'a}lez}, Antonio {Fuentes}, and John {Wardle}.
\newblock {Reading M87's DNA: A Double Helix Revealing a Large-scale Helical Magnetic Field}.
\newblock {\em ApJL}, 923(1):L5, December 2021.

\bibitem{2021ApJ...911L..11E}
{EHT MWL Science Working Group}, J.~C. {Algaba}, J.~{Anczarski}, K.~{Asada}, M.~{Balokovi{\'c}}, S.~{Chandra}, Y.~Z. {Cui}, A.~D. {Falcone}, M.~{Giroletti}, C.~{Goddi}, and et~al.
\newblock {Broadband Multi-wavelength Properties of M87 during the 2017 Event Horizon Telescope Campaign}.
\newblock {\em ApJL}, 911(1):L11, April 2021.

\bibitem{2021ApJ...909...80B}
Eduardo {Ba{\~n}ados}, Chiara {Mazzucchelli}, Emmanuel {Momjian}, Anna-Christina {Eilers}, Feige {Wang}, Jan-Torge {Schindler}, Thomas {Connor}, Irham~Taufik {Andika}, Aaron~J. {Barth}, Chris {Carilli}, Frederick~B. {Davies}, Roberto {Decarli}, Xiaohui {Fan}, Emanuele~Paolo {Farina}, Joseph~F. {Hennawi}, Antonio {Pensabene}, Daniel {Stern}, Bram~P. {Venemans}, Lukas {Wenzl}, and Jinyi {Yang}.
\newblock {The Discovery of a Highly Accreting, Radio-loud Quasar at z = 6.82}.
\newblock {\em ApJ}, 909(1):80, March 2021.

\bibitem{2021AJ....161..207M}
Emmanuel {Momjian}, Eduardo {Ba{\~n}ados}, Christopher~L. {Carilli}, Fabian {Walter}, and Chiara {Mazzucchelli}.
\newblock {Resolving the Radio Emission from the Quasar P172+18 at z = 6.82}.
\newblock {\em Astrophys.J.}, 161(5):207, 2021.

\bibitem{2021AN....342.1117L}
Matthew~L. {Lister}, Dan~C. {Homan}, Yuri~Y. {Kovalev}, Soham {Mandal}, Alexander~B. {Pushkarev}, and Aneta {Siemiginowska}.
\newblock {TXS 0128+554: A young gamma‑ray emitting active galactic nucleus with episodic jet activity}.
\newblock {\em Astronomische Nachrichten}, 342(1117):1117--1120, November 2021.

\bibitem{JetPress1}
Vla reveals double-helix structure in massive galaxy’s jet.
\newblock \url{https://public.nrao.edu/news/helix-structure-in-jet/}, 2021.

\bibitem{JetPress2}
Multi-wavelength observations reveal impact of black hole on m87 galaxy.
\newblock \url{https://public.nrao.edu/news/m87-black-hole-multi-wavelength-eht/}, 2021.

\bibitem{JetPress3}
Most distant cosmic jet providing clues about early universe.
\newblock \url{https://public.nrao.edu/news/distant-cosmic-jet/}, 2021.

\bibitem{JetPress4}
Vla sky survey reveals newborn jets in distant galaxies.
\newblock \url{https://public.nrao.edu/news/newborn-jets-distant-galaxies/}, 2020.

\bibitem{JetPress5}
A galaxy’s stop-and-start young radio jets.
\newblock \url{https://public.nrao.edu/news/stop-start-jets/}, 2020.

\bibitem{JetPress6}
Black hole’s tug on space pulls fast-moving jets in rapid wobble.
\newblock \url{https://public.nrao.edu/news/black-hole-jets-rapid-wobble/}, 2019.

\bibitem{JetPress7}
Vla discovers powerful jet coming from 'wrong' kind of star.
\newblock \url{https://public.nrao.edu/news/jet-from-wrong-kind-of-star/}, 2018.

\bibitem{1999MNRAS.309L...7G}
Gabriele {Ghisellini} and Davide {Lazzati}.
\newblock {Polarization light curves and position angle variation of beamed gamma-ray bursts}.
\newblock {\em MNRAS}, 309(1):L7--L11, October 1999.

\bibitem{2018MNRAS.478.4128G}
Ramandeep {Gill} and Jonathan {Granot}.
\newblock {Afterglow imaging and polarization of misaligned structured GRB jets and cocoons: breaking the degeneracy in GRB 170817A}.
\newblock {\em MNRAS}, 478(3):4128--4141, August 2018.

\bibitem{MWPress}
New study reveals previously unseen star formation in milky way.
\newblock \url{https://public.nrao.edu/news/new-study-star-formation-milky-way/}, 2022.

\bibitem{NurseryPress}
New look at a bright stellar nursery.
\newblock \url{https://public.nrao.edu/news/new-look-bright-stellar-nursery/}, 2021.

\bibitem{DustPress}
Astronomers find elusive target hiding behind dust.
\newblock \url{https://public.nrao.edu/news/target-hiding-behind-dust/}, 2020.

\bibitem{2013MNRAS.429..987L}
S.~N. {Longmore}, J.~{Bally}, L.~{Testi}, C.~R. {Purcell}, A.~J. {Walsh}, E.~{Bressert}, M.~{Pestalozzi}, S.~{Molinari}, J.~{Ott}, L.~{Cortese}, C.~{Battersby}, N.~{Murray}, E.~{Lee}, J.~M.~D. {Kruijssen}, E.~{Schisano}, and D.~{Elia}.
\newblock {Variations in the Galactic star formation rate and density thresholds for star formation}.
\newblock {\em MNRAS}, 429(2):987--1000, February 2013.

\bibitem{2020AJ....160..234D}
C.~G. {De Pree}, D.~J. {Wilner}, L.~E. {Kristensen}, R.~{Galv{\'a}n-Madrid}, W.~M. {Goss}, R.~S. {Klessen}, M.~M. {Mac Low}, T.~{Peters}, A.~{Robinson}, S.~{Sloman}, and M.~{Rao}.
\newblock {Time-variable Radio Recombination Line Emission in W49A}.
\newblock {\em Astrophys.J.}, 160(5):234, November 2020.

\bibitem{2020ApJ...896L...3D}
Marta {De Simone}, Cecilia {Ceccarelli}, Claudio {Codella}, Brian~E. {Svoboda}, Claire {Chandler}, Mathilde {Bouvier}, Satoshi {Yamamoto}, Nami {Sakai}, Paola {Caselli}, Cecile {Favre}, Laurent {Loinard}, Bertrand {Lefloch}, Hauyu~Baobab {Liu}, Ana {L{\'o}pez-Sepulcre}, Jaime~E. {Pineda}, Vianney {Taquet}, and Leonardo {Testi}.
\newblock {Hot Corinos Chemical Diversity: Myth or Reality?}
\newblock {\em ApJL}, 896(1):L3, June 2020.

\bibitem{2021A&A...651A..85B}
A.~{Brunthaler}, K.~M. {Menten}, S.~A. {Dzib}, W.~D. {Cotton}, F.~{Wyrowski}, R.~{Dokara}, Y.~{Gong}, S.~N.~X. {Medina}, P.~{M{\"u}ller}, H.~{Nguyen}, G.~N. {Ortiz-Le{\'o}n}, W.~{Reich}, M.~R. {Rugel}, J.~S. {Urquhart}, B.~{Winkel}, A.~Y. {Yang}, H.~{Beuther}, S.~{Billington}, C.~{Carrasco-Gonzalez}, T.~{Csengeri}, C.~{Murugeshan}, J.~D. {Pandian}, and N.~{Roy}.
\newblock {A global view on star formation: The GLOSTAR Galactic plane survey. I. Overview and first results for the Galactic longitude range 28{\textdegree} < l < 36{\textdegree}}.
\newblock {\em A\&A}, 651:A85, July 2021.

\bibitem{2019ApJS..244...35L}
Xing {Lu}, Elisabeth A.~C. {Mills}, Adam {Ginsburg}, Daniel~L. {Walker}, Ashley~T. {Barnes}, Natalie {Butterfield}, Jonathan~D. {Henshaw}, Cara {Battersby}, J.~M.~Diederik {Kruijssen}, Steven~N. {Longmore}, Qizhou {Zhang}, John {Bally}, Jens {Kauffmann}, J{\"u}rgen {Ott}, Matthew {Rickert}, and Ke~{Wang}.
\newblock {A Census of Early-phase High-mass Star Formation in the Central Molecular Zone}.
\newblock {\em ApJs}, 244(2):35, October 2019.

\bibitem{2021ApJ...906...24D}
Sergio~A. {Dzib}, Jan {Forbrich}, Mark~J. {Reid}, and Karl~M. {Menten}.
\newblock {A VLBA Survey of Radio Stars in the Orion Nebula Cluster. II.~Astrometry}.
\newblock {\em ApJ}, 906(1):24, January 2021.

\bibitem{2017ApJ...834..139D}
Sergio~A. {Dzib}, Laurent {Loinard}, Luis~F. {Rodr{\'\i}guez}, Laura {G{\'o}mez}, Jan {Forbrich}, Karl~M. {Menten}, Marina~A. {Kounkel}, Amy~J. {Mioduszewski}, Lee {Hartmann}, John~J. {Tobin}, and Juana~L. {Rivera}.
\newblock {Radio Measurements of the Stellar Proper Motions in the Core of the Orion Nebula Cluster}.
\newblock {\em ApJ}, 834(2):139, January 2017.

\bibitem{2014ApJ...791..124G}
Gilberto~C. {G{\'o}mez} and Enrique {V{\'a}zquez-Semadeni}.
\newblock {Filaments in Simulations of Molecular Cloud Formation}.
\newblock {\em ApJ}, 791(2):124, August 2014.

\bibitem{2012ApJ...748...16T}
John~J. {Tobin}, Lee {Hartmann}, Edwin {Bergin}, Hsin-Fang {Chiang}, Leslie~W. {Looney}, Claire~J. {Chandler}, S{\'e}bastien {Maret}, and Fabian {Heitsch}.
\newblock {Complex Structure in Class 0 Protostellar Envelopes. III. Velocity Gradients in Non-axisymmetric Envelopes, Infall, or Rotation?}
\newblock {\em ApJ}, 748(1):16, March 2012.

\bibitem{2019ApJ...882..103P}
Jaime~E. {Pineda}, Bo~{Zhao}, Anika {Schmiedeke}, Dominique~M. {Segura-Cox}, Paola {Caselli}, Philip~C. {Myers}, John~J. {Tobin}, and Michael {Dunham}.
\newblock {The Specific Angular Momentum Radial Profile in Dense Cores: Improved Initial Conditions for Disk Formation}.
\newblock {\em ApJ}, 882(2):103, September 2019.

\bibitem{2020A&A...644A.128S}
Inma {Sep{\'u}lveda}, Robert {Estalella}, Guillem {Anglada}, Rosario {L{\'o}pez}, Angels {Riera}, Gemma {Busquet}, Aina {Palau}, Jos{\'e}~M. {Torrelles}, and Luis~F. {Rodr{\'\i}guez}.
\newblock {VLA ammonia observations of L1287. Analysis of the Guitar core and two filaments}.
\newblock {\em A\&A}, 644:A128, December 2020.

\bibitem{2020ApJ...890..129K}
N.~{Karnath}, S.~T. {Megeath}, J.~J. {Tobin}, A.~{Stutz}, Z.~Y. {Li}, P.~{Sheehan}, N.~{Reynolds}, S.~{Sadavoy}, I.~W. {Stephens}, M.~{Osorio}, G.~{Anglada}, A.~K. {D{\'\i}az-Rodr{\'\i}guez}, and E.~{Cox}.
\newblock {Detection of Irregular, Submillimeter Opaque Structures in the Orion Molecular Clouds: Protostars within 10,000 yr of Formation?}
\newblock {\em ApJ}, 890(2):129, February 2020.

\bibitem{2020ApJ...890..130T}
John~J. {Tobin}, Patrick~D. {Sheehan}, S.~Thomas {Megeath}, Ana~Karla {D{\'\i}az-Rodr{\'\i}guez}, Stella S.~R. {Offner}, Nadia~M. {Murillo}, Merel L.~R. {van 't Hoff}, Ewine~F. {van Dishoeck}, Mayra {Osorio}, Guillem {Anglada}, Elise {Furlan}, Amelia~M. {Stutz}, Nickalas {Reynolds}, Nicole {Karnath}, William~J. {Fischer}, Magnus {Persson}, Leslie~W. {Looney}, Zhi-Yun {Li}, Ian {Stephens}, Claire~J. {Chandler}, Erin {Cox}, Michael~M. {Dunham}, {\L}ukasz {Tychoniec}, Mihkel {Kama}, Kaitlin {Kratter}, Marina {Kounkel}, Brian {Mazur}, Luke {Maud}, Lisa {Patel}, Laura {Perez}, Sarah~I. {Sadavoy}, Dominique {Segura-Cox}, Rajeeb {Sharma}, Brian {Stephenson}, Dan~M. {Watson}, and Friedrich {Wyrowski}.
\newblock {The VLA/ALMA Nascent Disk and Multiplicity (VANDAM) Survey of Orion Protostars. II. A Statistical Characterization of Class 0 and Class I Protostellar Disks}.
\newblock {\em ApJ}, 890(2):130, February 2020.

\bibitem{2022MNRAS.516..185K}
Atanu {Koley}.
\newblock {Studying the chemical and kinematical structures of dense cores TMC-1C, L1544, and TMC-1 in the Taurus molecular cloud using CCS and NH$_{3}$ observations}.
\newblock {\em MNRAS}, 516(1):185--196, October 2022.

\bibitem{PlanetBirthPress}
How newborn stars prepare for the birth of planets.
\newblock \url{https://public.nrao.edu/news/how-newborn-stars-prepare-for-the-birth-of-planets/}, 2020.

\bibitem{2012ApJ...746..174R}
Sarah~E. {Ragan}, Fabian {Heitsch}, Edwin~A. {Bergin}, and David {Wilner}.
\newblock {Very Large Array Observations of Ammonia in Infrared-dark Clouds. II. Internal Kinematics}.
\newblock {\em ApJ}, 746(2):174, February 2012.

\bibitem{2022MNRAS.516L..48K}
Atanu {Koley}, Nirupam {Roy}, Emmanuel {Momjian}, Anuj~P. {Sarma}, and Abhirup {Datta}.
\newblock {Magnetic field measurement in TMC-1C using 22.3 GHz CCS Zeeman splitting}.
\newblock {\em MNRAS}, 516(1):L48--L52, October 2022.

\bibitem{2018ApJ...856...39C}
M.~K. {Crosley} and R.~A. {Osten}.
\newblock {Constraining Stellar Coronal Mass Ejections through Multi-wavelength Analysis of the Active M Dwarf EQ Peg}.
\newblock {\em ApJ}, 856(1):39, March 2018.

\bibitem{2019ApJ...871..214V}
Jackie {Villadsen} and Gregg {Hallinan}.
\newblock {Ultra-wideband Detection of~22 Coherent Radio Bursts on M Dwarfs}.
\newblock {\em ApJ}, 871(2):214, February 2019.

\bibitem{2020ApJ...901...98L}
Megan~O. {Lewis}, Ylva~M. {Pihlstr{\"o}m}, Lor{\'a}nt~O. {Sjouwerman}, and Luis~Henry {Quiroga-Nu{\~n}ez}.
\newblock {Infrared Color Separation between Thin-shelled Oxygen-rich and Carbon-rich AGB Stars}.
\newblock {\em ApJ}, 901(2):98, October 2020.

\bibitem{2017PhDT........85V}
Jackie {Villadsen}.
\newblock {\em {The Search for Stellar Coronal Mass Ejections}}.
\newblock PhD thesis, California Institute of Technology, Division of Physics, Mathematics and Astronomy, January 2017.

\bibitem{2023ApJ...951L...9A}
Gabriella {Agazie}, Md~Faisal {Alam}, Akash {Anumarlapudi}, Anne~M. {Archibald}, Zaven {Arzoumanian}, Paul~T. {Baker}, Laura {Blecha}, Victoria {Bonidie}, Adam {Brazier}, Paul~R. {Brook}, Sarah {Burke-Spolaor}, Bence {B{\'e}csy}, Christopher {Chapman}, Maria {Charisi}, Shami {Chatterjee}, Tyler {Cohen}, James~M. {Cordes}, Neil~J. {Cornish}, Fronefield {Crawford}, H.~Thankful {Cromartie}, Kathryn {Crowter}, Megan~E. {Decesar}, Paul~B. {Demorest}, Timothy {Dolch}, Brendan {Drachler}, Elizabeth~C. {Ferrara}, William {Fiore}, Emmanuel {Fonseca}, Gabriel~E. {Freedman}, Nate {Garver-Daniels}, Peter~A. {Gentile}, Joseph {Glaser}, Deborah~C. {Good}, Kayhan {G{\"u}ltekin}, Jeffrey~S. {Hazboun}, Ross~J. {Jennings}, Cody {Jessup}, Aaron~D. {Johnson}, Megan~L. {Jones}, Andrew~R. {Kaiser}, David~L. {Kaplan}, Luke~Zoltan {Kelley}, Matthew {Kerr}, Joey~S. {Key}, Anastasia {Kuske}, Nima {Laal}, Michael~T. {Lam}, William~G. {Lamb}, T.~Joseph~W. {Lazio}, Natalia {Lewandowska}, Ye~{Lin}, Tingting {Liu}, Duncan~R. {Lorimer},
  Jing {Luo}, Ryan~S. {Lynch}, Chung-Pei {Ma}, Dustin~R. {Madison}, Kaleb {Maraccini}, Alexander {McEwen}, James~W. {McKee}, Maura~A. {McLaughlin}, Natasha {McMann}, Bradley~W. {Meyers}, Chiara M.~F. {Mingarelli}, Andrea {Mitridate}, Cherry {Ng}, David~J. {Nice}, Stella~Koch {Ocker}, Ken~D. {Olum}, Elisa {Panciu}, Timothy~T. {Pennucci}, Benetge B.~P. {Perera}, Nihan~S. {Pol}, Henri~A. {Radovan}, Scott~M. {Ransom}, Paul~S. {Ray}, Joseph~D. {Romano}, Laura {Salo}, Shashwat~C. {Sardesai}, Carl {Schmiedekamp}, Ann {Schmiedekamp}, Kai {Schmitz}, Brent~J. {Shapiro-Albert}, Xavier {Siemens}, Joseph {Simon}, Magdalena~S. {Siwek}, Ingrid~H. {Stairs}, Daniel~R. {Stinebring}, Kevin {Stovall}, Abhimanyu {Susobhanan}, Joseph~K. {Swiggum}, Stephen~R. {Taylor}, Jacob~E. {Turner}, Caner {Unal}, Michele {Vallisneri}, Sarah~J. {Vigeland}, Haley~M. {Wahl}, Qiaohong {Wang}, Caitlin~A. {Witt}, Olivia {Young}, and {Nanograv Collaboration}.
\newblock {The NANOGrav 15~yr Data Set: Observations and Timing of 68 Millisecond Pulsars}.
\newblock {\em ApJL}, 951(1):L9, July 2023.

\bibitem{cromartie2020}
H.~T. {Cromartie}, E.~{Fonseca}, S.~M. {Ransom}, P.~B. {Demorest}, Z.~{Arzoumanian}, H.~{Blumer}, P.~R. {Brook}, M.~E. {DeCesar}, T.~{Dolch}, J.~A. {Ellis}, R.~D. {Ferdman}, E.~C. {Ferrara}, N.~{Garver-Daniels}, P.~A. {Gentile}, M.~L. {Jones}, M.~T. {Lam}, D.~R. {Lorimer}, R.~S. {Lynch}, M.~A. {McLaughlin}, C.~{Ng}, D.~J. {Nice}, T.~T. {Pennucci}, R.~{Spiewak}, I.~H. {Stairs}, K.~{Stovall}, J.~K. {Swiggum}, and W.~W. {Zhu}.
\newblock {Relativistic Shapiro delay measurements of an extremely massive millisecond pulsar}.
\newblock {\em Nature Astronomy}, 4:72--76, January 2020.

\bibitem{WanderingBHPress}
Astronomers find wandering massive black holes in dwarf galaxies.
\newblock \url{https://public.nrao.edu/news/wandering-black-holes-dwarf-galaxies/}, 2020.

\bibitem{2020ApJ...888...36R}
Amy~E. {Reines}, James~J. {Condon}, Jeremy {Darling}, and Jenny~E. {Greene}.
\newblock {A New Sample of (Wandering) Massive Black Holes in Dwarf Galaxies from High-resolution Radio Observations}.
\newblock {\em ApJ}, 888(1):36, January 2020.

\bibitem{2022ApJ...941...43Y}
Xiaolong {Yang}, Prashanth {Mohan}, Jun {Yang}, Luis~C. {Ho}, J.~N.~H.~S. {Aditya}, Shaohua {Zhang}, Sumit {Jaiswal}, and Xiaofeng {Yang}.
\newblock {Radio Observations of Four Active Galactic Nuclei Hosting Intermediate-mass Black Hole Candidates: Studying the Outflow Activity and Evolution}.
\newblock {\em ApJ}, 941(1):43, December 2022.

\bibitem{2022ApJ...933..160S}
Andrew~J. {Sargent}, Megan~C. {Johnson}, Amy~E. {Reines}, Nathan~J. {Secrest}, Alexander~J. {van~der Horst}, Phil~J. {Cigan}, Jeremy {Darling}, and Jenny~E. {Greene}.
\newblock {Wandering Black Hole Candidates in Dwarf Galaxies at VLBI Resolution}.
\newblock {\em ApJ}, 933(2):160, July 2022.

\bibitem{2020ARA&A..58..257G}
Jenny~E. {Greene}, Jay {Strader}, and Luis~C. {Ho}.
\newblock {Intermediate-Mass Black Holes}.
\newblock {\em ARA\&A}, 58:257--312, August 2020.

\bibitem{2023ApJ...958...29C}
Yu-Ching {Chen}, Xin {Liu}, Joseph {Lazio}, Peter {Breiding}, Sarah {Burke-Spolaor}, Hsiang-Chih {Hwang}, Yue {Shen}, and Nadia~L. {Zakamska}.
\newblock {Varstrometry for Off-nucleus and Dual Sub-kiloparsec Active Galactic Nuclei (VODKA): Very Long Baseline Array Searches for Dual or Off-nucleus Quasars and Small-scale Jets}.
\newblock {\em ApJ}, 958(1):29, November 2023.

\bibitem{wang2021}
Feige {Wang}, Jinyi {Yang}, Xiaohui {Fan}, Joseph~F. {Hennawi}, Aaron~J. {Barth}, Eduardo {Banados}, Fuyan {Bian}, Konstantina {Boutsia}, Thomas {Connor}, Frederick~B. {Davies}, Roberto {Decarli}, Anna-Christina {Eilers}, Emanuele~Paolo {Farina}, Richard {Green}, Linhua {Jiang}, Jiang-Tao {Li}, Chiara {Mazzucchelli}, Riccardo {Nanni}, Jan-Torge {Schindler}, Bram {Venemans}, Fabian {Walter}, Xue-Bing {Wu}, and Minghao {Yue}.
\newblock {A Luminous Quasar at Redshift 7.642}.
\newblock {\em ApJL}, 907(1):L1, January 2021.

\bibitem{endsley2022}
Ryan {Endsley}, Daniel~P. {Stark}, Xiaohui {Fan}, Renske {Smit}, Feige {Wang}, Jinyi {Yang}, Kevin {Hainline}, Jianwei {Lyu}, Rychard {Bouwens}, and Sander {Schouws}.
\newblock {Radio and far-IR emission associated with a massive star-forming galaxy candidate at $z \simeq 6.8$: a radio-loud AGN in the reionization era?}
\newblock {\em MNRAS}, 512(3):4248--4261, 2022.

\bibitem{2017AJ....153..132A}
Trisha {Ashley}, Caroline~E. {Simpson}, Bruce~G. {Elmegreen}, Megan {Johnson}, and Nau~Raj {Pokhrel}.
\newblock {The HI Chronicles of LITTLE THINGS BCDs. III. Gas Clouds in and around Mrk 178, VII Zw 403, and NGC 3738}.
\newblock {\em Astrophys.J.}, 153(3):132, March 2017.

\bibitem{2018MNRAS.473.5248O}
Ewan {O'Sullivan}, Konstantinos {Kolokythas}, Nimisha~G. {Kantharia}, Somak {Raychaudhury}, Laurence~P. {David}, and Jan~M. {Vrtilek}.
\newblock {The origin of the X-ray, radio and H I structures in the NGC 5903 galaxy group}.
\newblock {\em MNRAS}, 473(4):5248--5266, February 2018.

\bibitem{2023A&A...670A..21J}
M.~G. {Jones}, L.~{Verdes-Montenegro}, J.~{Moldon}, A.~{Damas Segovia}, S.~{Borthakur}, S.~{Luna}, M.~{Yun}, A.~{del Olmo}, J.~{Perea}, J.~{Cannon}, D.~{Lopez Gutierrez}, M.~{Cluver}, J.~{Garrido}, and S.~{Sanchez}.
\newblock {Disturbed, diffuse, or just missing? A global study of the H I content of Hickson compact groups}.
\newblock {\em A\&A}, 670:A21, February 2023.

\bibitem{2013ApJ...779..173N}
Kristina {Nyland}, Katherine {Alatalo}, J.~M. {Wrobel}, Lisa~M. {Young}, Raffaella {Morganti}, Timothy~A. {Davis}, P.~T. {de Zeeuw}, Susana {Deustua}, and Martin {Bureau}.
\newblock {Detection of a High Brightness Temperature Radio Core in the Active-galactic-nucleus-driven Molecular Outflow Candidate NGC 1266}.
\newblock {\em ApJ}, 779(2):173, December 2013.

\bibitem{2018A&A...617A..38S}
R.~{Schulz}, R.~{Morganti}, K.~{Nyland}, Z.~{Paragi}, E.~K. {Mahony}, and T.~{Oosterloo}.
\newblock {Mapping the neutral atomic hydrogen gas outflow in the restarted radio galaxy 3C 236}.
\newblock {\em A\&A}, 617:A38, September 2018.

\bibitem{CosmicRayPress}
Vla finds cosmic rays driving galaxy’s winds.
\newblock \url{https://public.nrao.edu/news/cosmic-rays-galaxys-winds/}, 2022.

\bibitem{RopePress}
Giant magnetic ropes in a galaxy’s halo.
\newblock \url{https://public.nrao.edu/news/giant-magnetic-ropes/}, 2019.

\bibitem{2015AJ....150...81W}
Theresa {Wiegert}, Judith {Irwin}, Arpad {Miskolczi}, Philip {Schmidt}, Silvia~Carolina {Mora}, Ancor {Damas-Segovia}, Yelena {Stein}, Jayanne {English}, Richard~J. {Rand}, Isaiah {Santistevan}, Rene {Walterbos}, Marita {Krause}, Rainer {Beck}, Ralf-J{\"u}rgen {Dettmar}, Amanda {Kepley}, Marek {Wezgowiec}, Q.~Daniel {Wang}, George {Heald}, Jiangtao {Li}, Stephen {MacGregor}, Megan {Johnson}, A.~W. {Strong}, Amanda {DeSouza}, and Troy~A. {Porter}.
\newblock {CHANG-ES. IV. Radio Continuum Emission of 35 Edge-on Galaxies Observed with the Karl G. Jansky Very Large Array in D Configuration{\textemdash}Data Release 1}.
\newblock {\em Astrophys.J.}, 150(3):81, September 2015.

\bibitem{2019MNRAS.487.1498W}
Alex {Woodfinden}, R.~N. {Henriksen}, Judith {Irwin}, and Silvia~Carolina {Mora-Partiarroyo}.
\newblock {Evolving galactic dynamos and fits to the reversing rotation measures in the halo of NGC 4631}.
\newblock {\em MNRAS}, 487(2):1498--1516, August 2019.

\bibitem{2022MNRAS.517.2990T}
F.~S. {Tabatabaei}, W.~{Cotton}, E.~{Schinnerer}, R.~{Beck}, A.~{Brunthaler}, K.~M. {Menten}, J.~{Braine}, E.~{Corbelli}, C.~{Kramer}, J.~E. {Beckman}, J.~H. {Knapen}, R.~{Paladino}, E.~{Koch}, and A.~{Camps Fari{\~n}a}.
\newblock {Cloud-scale radio surveys of star formation and feedback in Triangulum Galaxy M 33: VLA observations}.
\newblock {\em MNRAS}, 517(2):2990--3007, December 2022.

\bibitem{2024Galax..12...22I}
Judith {Irwin}, Rainer {Beck}, Tanden {Cook}, Ralf-J{\"u}rgen {Dettmar}, Jayanne {English}, Volker {Heesen}, Richard {Henriksen}, Yan {Jiang}, Jiang-Tao {Li}, Li-Yuan {Lu}, Crystal {Mele}, Ancla {M{\"u}ller}, Eric {Murphy}, Troy {Porter}, Richard {Rand}, Nathan {Skeggs}, Michael {Stein}, Yelena {Stein}, Jeroen {Stil}, Andrew {Strong}, Rene {Walterbos}, Q.~Daniel {Wang}, Theresa {Wiegert}, and Yang {Yang}.
\newblock {CHANG-ES~XXXI{\textemdash}A Decade of CHANG-ES: What We Have Learned from Radio Observations of Edge-on Galaxies}.
\newblock {\em Galaxies}, 12(3):22, 2024.

\bibitem{DoubleHelixPress}
Vla reveals double-helix structure in massive galaxy’s jet.
\newblock \url{https://public.nrao.edu/news/helix-structure-in-jet/}, 2021.

\bibitem{StormPress}
Vla finds unexpected storm at galaxy’s core.
\newblock \url{https://public.nrao.edu/news/vla-finds-galaxy-storm/}, 2015.

\bibitem{BlowingPress}
Powerful jets blowing material out of galaxy.
\newblock \url{https://public.nrao.edu/news/powerful-jets-blow-material-out-of-galaxy/}, 2013.

\bibitem{QuartetPress}
Astronomers spy quartet of cavities from giant black holes.
\newblock \url{https://www.nasa.gov/image-article/astronomers-spy-quartet-of-cavities-from-giant-black-holes/}, 2021.

\bibitem{WeakBHPress}
A weakened black hole allows its galaxy to awaken.
\newblock \url{https://public.nrao.edu/news/weakened-black-hole/}, 2019.

\bibitem{2019MNRAS.484.3376L}
Wenhao {Liu}, Ming {Sun}, Paul {Nulsen}, Tracy {Clarke}, Craig {Sarazin}, William {Forman}, Massimo {Gaspari}, Simona {Giacintucci}, Dharam~Vir {Lal}, and Tim {Edge}.
\newblock {AGN feedback in galaxy group 3C 88: cavities, shock, and jet reorientation}.
\newblock {\em MNRAS}, 484(3):3376--3392, April 2019.

\bibitem{2023A&A...670A..23U}
F.~{Ubertosi}, M.~{Gitti}, and F.~{Brighenti}.
\newblock {Chasing ICM cooling and AGN feedback from the macro to the meso scales in the galaxy cluster ZwCl 235}.
\newblock {\em A\&A}, 670:A23, February 2023.

\bibitem{1984Natur.310..557Y}
F.~{Yusef-Zadeh}, M.~{Morris}, and D.~{Chance}.
\newblock {Large, highly organized radio structures near the galactic centre}.
\newblock {\em Nature}, 310(5978):557--561, August 1984.

\bibitem{1985AJ.....90.2511M}
M.~{Morris} and F.~{Yusef-Zadeh}.
\newblock {Unusual threads of radio emission near the Galactic Center.}
\newblock {\em AJ}, 90:2511--2513, December 1985.

\bibitem{2019MNRAS.482.5349R}
Matthew {Rickert}, F.~{Yusef-Zadeh}, and J.~{Ott}.
\newblock {A 6.7~GHz methanol maser survey of the central molecular zone}.
\newblock {\em MNRAS}, 482(4):5349--5361, February 2019.

\bibitem{2022ApJ...936..186B}
Natalie~O. {Butterfield}, Cornelia~C. {Lang}, Adam {Ginsburg}, Mark~R. {Morris}, J{\"u}rgen {Ott}, and Dominic~A. {Ludovici}.
\newblock {Evidence for an Interaction between the Galactic Center Clouds M0.10$-$0.08 and M0.11$-$0.11}.
\newblock {\em ApJ}, 936(2):186, September 2022.

\bibitem{2022A&A...666A..31M}
F.~{Meng}, {\'A}.~{S{\'a}nchez-Monge}, P.~{Schilke}, A.~{Ginsburg}, C.~{DePree}, N.~{Budaiev}, D.~{Jeff}, A.~{Schmiedeke}, A.~{Schw{\"o}rer}, V.~S. {Veena}, and Th. {M{\"o}ller}.
\newblock {The physical and chemical structure of Sagittarius B2. VI.~UCH\textsc{ii} regions in Sgr~B2}.
\newblock {\em A\&A}, 666:A31, October 2022.

\bibitem{2023A&A...680A..43L}
Y.~{Lin}, S.~{Spezzano}, J.~E. {Pineda}, J.~{Harju}, A.~{Schmiedeke}, S.~{Jiao}, H.~B. {Liu}, and P.~{Caselli}.
\newblock {Initial conditions of star formation at {\ensuremath{\lesssim}}2000~au: Physical structure and NH$_{3}$ depletion of three early-stage cores}.
\newblock {\em A\&A}, 680:A43, December 2023.

\bibitem{2021NewAR..9301630B}
Aaron {Bryant} and Alfred {Krabbe}.
\newblock {The episodic and multiscale Galactic Centre}.
\newblock {\em NewAR}, 93:101630, December 2021.

\bibitem{2023ApJ...951L..11A}
Adeela {Afzal}, Gabriella {Agazie}, Akash {Anumarlapudi}, Anne~M. {Archibald}, Zaven {Arzoumanian}, Paul~T. {Baker}, Bence {B{\'{e}}csy}, Jose~Juan {Blanco-Pillado}, Laura {Blecha}, Kimberly~K. {Boddy}, Adam {Brazier}, Paul~R. {Brook}, Sarah {Burke-Spolaor}, Rand {Burnette}, Robin {Case}, Maria {Charisi}, Shami {Chatterjee}, Katerina {Chatziioannou}, Belinda~D. {Cheeseboro}, Siyuan {Chen}, Tyler {Cohen}, James~M. {Cordes}, Neil~J. {Cornish}, Fronefield {Crawford}, H.~Thankful {Cromartie}, Kathryn {Crowter}, Curt~J. {Cutler}, Megan~E. {Decesar}, Dallas {Degan}, Paul~B. {Demorest}, Heling {Deng}, Timothy {Dolch}, Brendan {Drachler}, Richard {von Eckardstein}, Elizabeth~C. {Ferrara}, William {Fiore}, Emmanuel {Fonseca}, Gabriel~E. {Freedman}, Nate {Garver-Daniels}, Peter~A. {Gentile}, Kyle~A. {Gersbach}, Joseph {Glaser}, Deborah~C. {Good}, Lydia {Guertin}, Kayhan {G{\"u}ltekin}, Jeffrey~S. {Hazboun}, Sophie {Hourihane}, Kristina {Islo}, Ross~J. {Jennings}, Aaron~D. {Johnson}, Megan~L. {Jones}, Andrew~R.
  {Kaiser}, David~L. {Kaplan}, Luke~Zoltan {Kelley}, Matthew {Kerr}, Joey~S. {Key}, Nima {Laal}, Michael~T. {Lam}, William~G. {Lamb}, T.~Joseph~W. {Lazio}, Vincent S.~H. {Lee}, Natalia {Lewandowska}, Rafael~R. {Lino Dos Santos}, Tyson~B. {Littenberg}, Tingting {Liu}, Duncan~R. {Lorimer}, Jing {Luo}, Ryan~S. {Lynch}, Chung-Pei {Ma}, Dustin~R. {Madison}, Alexander {McEwen}, James~W. {McKee}, Maura~A. {McLaughlin}, Natasha {McMann}, Bradley~W. {Meyers}, Patrick~M. {Meyers}, Chiara M.~F. {Mingarelli}, Andrea {Mitridate}, Jonathan {Nay}, Priyamvada {Natarajan}, Cherry {Ng}, David~J. {Nice}, Stella~Koch {Ocker}, Ken~D. {Olum}, Timothy~T. {Pennucci}, Benetge B.~P. {Perera}, Polina {Petrov}, Nihan~S. {Pol}, Henri~A. {Radovan}, Scott~M. {Ransom}, Paul~S. {Ray}, Joseph~D. {Romano}, Shashwat~C. {Sardesai}, Ann {Schmiedekamp}, Carl {Schmiedekamp}, Kai {Schmitz}, Tobias {Schr{\"o}der}, Levi {Schult}, Brent~J. {Shapiro-Albert}, Xavier {Siemens}, Joseph {Simon}, Magdalena~S. {Siwek}, Ingrid~H. {Stairs}, Daniel~R.
  {Stinebring}, Kevin {Stovall}, Peter {Stratmann}, Jerry~P. {Sun}, Abhimanyu {Susobhanan}, Joseph~K. {Swiggum}, Jacob {Taylor}, Stephen~R. {Taylor}, Tanner {Trickle}, Jacob~E. {Turner}, Caner {Unal}, Michele {Vallisneri}, Sonali {Verma}, Sarah~J. {Vigeland}, Haley~M. {Wahl}, Qiaohong {Wang}, Caitlin~A. {Witt}, David {Wright}, Olivia {Young}, Kathryn~M. {Zurek}, and {Nanograv Collaboration}.
\newblock {The NANOGrav 15~yr Data Set: Search for Signals from New Physics}.
\newblock {\em ApJL}, 951(1):L11, July 2023.

\bibitem{chakrabarti2021}
Sukanya {Chakrabarti}, Philip {Chang}, Michael~T. {Lam}, Sarah~J. {Vigeland}, and Alice~C. {Quillen}.
\newblock {A Measurement of the Galactic Plane Mass Density from Binary Pulsar Accelerations}.
\newblock {\em ApJL}, 907(2):L26, February 2021.

\bibitem{2017MNRAS.468.2526B}
D.~{Bhakta}, J.~S. {Deneva}, D.~A. {Frail}, F.~{de Gasperin}, H.~T. {Intema}, P.~{Jagannathan}, and K.~P. {Mooley}.
\newblock {Searching for pulsars associated with the Fermi GeV excess}.
\newblock {\em MNRAS}, 468(3):2526--2531, July 2017.

\bibitem{2023ApJ...951L...8A}
Gabriella {Agazie}, Akash {Anumarlapudi}, Anne~M. {Archibald}, Zaven {Arzoumanian}, Paul~T. {Baker}, Bence {B{\'{e}}csy}, Laura {Blecha}, Adam {Brazier}, Paul~R. {Brook}, Sarah {Burke-Spolaor}, Rand {Burnette}, Robin {Case}, Maria {Charisi}, Shami {Chatterjee}, Katerina {Chatziioannou}, Belinda~D. {Cheeseboro}, Siyuan {Chen}, Tyler {Cohen}, James~M. {Cordes}, Neil~J. {Cornish}, Fronefield {Crawford}, H.~Thankful {Cromartie}, Kathryn {Crowter}, Curt~J. {Cutler}, Megan~E. {Decesar}, Dallas {Degan}, Paul~B. {Demorest}, Heling {Deng}, Timothy {Dolch}, Brendan {Drachler}, Justin~A. {Ellis}, Elizabeth~C. {Ferrara}, William {Fiore}, Emmanuel {Fonseca}, Gabriel~E. {Freedman}, Nate {Garver-Daniels}, Peter~A. {Gentile}, Kyle~A. {Gersbach}, Joseph {Glaser}, Deborah~C. {Good}, Kayhan {G{\"u}ltekin}, Jeffrey~S. {Hazboun}, Sophie {Hourihane}, Kristina {Islo}, Ross~J. {Jennings}, Aaron~D. {Johnson}, Megan~L. {Jones}, Andrew~R. {Kaiser}, David~L. {Kaplan}, Luke~Zoltan {Kelley}, Matthew {Kerr}, Joey~S. {Key}, Tonia~C.
  {Klein}, Nima {Laal}, Michael~T. {Lam}, William~G. {Lamb}, T.~Joseph~W. {Lazio}, Natalia {Lewandowska}, Tyson~B. {Littenberg}, Tingting {Liu}, Andrea {Lommen}, Duncan~R. {Lorimer}, Jing {Luo}, Ryan~S. {Lynch}, Chung-Pei {Ma}, Dustin~R. {Madison}, Margaret~A. {Mattson}, Alexander {McEwen}, James~W. {McKee}, Maura~A. {McLaughlin}, Natasha {McMann}, Bradley~W. {Meyers}, Patrick~M. {Meyers}, Chiara M.~F. {Mingarelli}, Andrea {Mitridate}, Priyamvada {Natarajan}, Cherry {Ng}, David~J. {Nice}, Stella~Koch {Ocker}, Ken~D. {Olum}, Timothy~T. {Pennucci}, Benetge B.~P. {Perera}, Polina {Petrov}, Nihan~S. {Pol}, Henri~A. {Radovan}, Scott~M. {Ransom}, Paul~S. {Ray}, Joseph~D. {Romano}, Shashwat~C. {Sardesai}, Ann {Schmiedekamp}, Carl {Schmiedekamp}, Kai {Schmitz}, Levi {Schult}, Brent~J. {Shapiro-Albert}, Xavier {Siemens}, Joseph {Simon}, Magdalena~S. {Siwek}, Ingrid~H. {Stairs}, Daniel~R. {Stinebring}, Kevin {Stovall}, Jerry~P. {Sun}, Abhimanyu {Susobhanan}, Joseph~K. {Swiggum}, Jacob {Taylor}, Stephen~R. {Taylor},
  Jacob~E. {Turner}, Caner {Unal}, Michele {Vallisneri}, Rutger {van Haasteren}, Sarah~J. {Vigeland}, Haley~M. {Wahl}, Qiaohong {Wang}, Caitlin~A. {Witt}, Olivia {Young}, and {Nanograv Collaboration}.
\newblock {The NANOGrav 15~yr Data Set: Evidence for a Gravitational-wave Background}.
\newblock {\em ApJL}, 951(1):L8, July 2023.

\bibitem{hotokezaka2019}
K.~{Hotokezaka}, E.~{Nakar}, O.~{Gottlieb}, S.~{Nissanke}, K.~{Masuda}, G.~{Hallinan}, K.~P. {Mooley}, and A.~T. {Deller}.
\newblock {A Hubble constant measurement from superluminal motion of the jet in GW170817}.
\newblock {\em Nature Astronomy}, 3:940--944, July 2019.

\bibitem{SpaceRadar}
Cross-disciplinary deep space radar needs study.
\newblock \url{https://www.nasa.gov/wp-content/uploads/2023/10/atr-2023-01267.pdf}, 2023.

\bibitem{DefensePress}
Future of earth’s defense is ground-based planetary radar.
\newblock \url{https://public.nrao.edu/news/planetary-scientists-need-radar/}, 2022.

\bibitem{TychoPress}
Moon’s tycho crater revealed in intricate detail.
\newblock \url{https://public.nrao.edu/news/radar-tycho-crater-intricate-detail/}, 2021.

\bibitem{2016A&A...588A.112G}
G.~{Guidi}, M.~{Tazzari}, L.~{Testi}, I.~{de Gregorio-Monsalvo}, C.~J. {Chandler}, L.~{P{\'e}rez}, A.~{Isella}, A.~{Natta}, S.~{Ortolani}, Th. {Henning}, S.~{Corder}, H.~{Linz}, S.~{Andrews}, D.~{Wilner}, L.~{Ricci}, J.~{Carpenter}, A.~{Sargent}, L.~{Mundy}, S.~{Storm}, N.~{Calvet}, C.~{Dullemond}, J.~{Greaves}, J.~{Lazio}, A.~{Deller}, and W.~{Kwon}.
\newblock {Dust properties across the CO snowline in the HD 163296 disk from ALMA and VLA observations}.
\newblock {\em A\&A}, 588:A112, April 2016.

\bibitem{2016ApJ...821L..16C}
Carlos {Carrasco-Gonz{\'a}lez}, Thomas {Henning}, Claire~J. {Chandler}, Hendrik {Linz}, Laura {P{\'e}rez}, Luis~F. {Rodr{\'\i}guez}, Roberto {Galv{\'a}n-Madrid}, Guillem {Anglada}, Til {Birnstiel}, Roy {van Boekel}, Mario {Flock}, Hubert {Klahr}, Enrique {Macias}, Karl {Menten}, Mayra {Osorio}, Leonardo {Testi}, Jos{\'e}~M. {Torrelles}, and Zhaohuan {Zhu}.
\newblock {The VLA View of the HL Tau Disk: Disk Mass, Grain Evolution, and Early Planet Formation}.
\newblock {\em ApJL}, 821(1):L16, April 2016.

\bibitem{2019ApJ...883...71C}
Carlos {Carrasco-Gonz{\'a}lez}, Anibal {Sierra}, Mario {Flock}, Zhaohuan {Zhu}, Thomas {Henning}, Claire {Chandler}, Roberto {Galv{\'a}n-Madrid}, Enrique {Mac{\'\i}as}, Guillem {Anglada}, Hendrik {Linz}, Mayra {Osorio}, Luis~F. {Rodr{\'\i}guez}, Leonardo {Testi}, Jos{\'e}~M. {Torrelles}, Laura {P{\'e}rez}, and Yao {Liu}.
\newblock {The Radial Distribution of Dust Particles in the HL Tau Disk from ALMA and VLA Observations}.
\newblock {\em ApJ}, 883(1):71, September 2019.

\bibitem{depater23}
Imke {de Pater}, Edward~M. {Molter}, and Chris~M. {Moeckel}.
\newblock {A Review of Radio Observations of the Giant Planets: Probing the Composition, Structure, and Dynamics of Their Deep Atmospheres}.
\newblock {\em Remote Sensing}, 15(5):1313, February 2023.

\bibitem{JupyterPress}
La and alma study jupiter and io.
\newblock \url{https://public.nrao.edu/news/vla-and-alma-study-jupiter-and-io/}, 2022.

\bibitem{2019Icar..322..168D}
Imke {de Pater}, R.~J. {Sault}, Michael~H. {Wong}, Leigh~N. {Fletcher}, David {DeBoer}, and Bryan {Butler}.
\newblock {Jupiter's ammonia distribution derived from VLA maps at 3-37 GHz}.
\newblock {\em Icar}, 322:168--191, April 2019.

\bibitem{2022PhDT........15M}
Edward~Mischel {Molter}.
\newblock {\em {Cloud Formation and Circulation in Planetary Tropospheres from Remote-Sensing Data}}.
\newblock PhD thesis, University of California, Berkeley, January 2022.

\bibitem{2023PSJ.....4...25M}
Chris {Moeckel}, Imke {de Pater}, and David {DeBoer}.
\newblock {Ammonia Abundance Derived from Juno MWR and VLA Observations of Jupiter}.
\newblock {\em PSJ}, 4(2):25, February 2023.

\bibitem{2023GeoRL..5002872A}
Alex {Akins}, Mark {Hofstadter}, Bryan {Butler}, A.~James {Friedson}, Edward {Molter}, Marzia {Parisi}, and Imke {de Pater}.
\newblock {Evidence of a Polar Cyclone on Uranus From VLA Observations}.
\newblock {\em GeoRL}, 50(10):e2023GL102872, 2023.

\bibitem{2017PhDT.........3Z}
Zhimeng {Zhang}.
\newblock {\em {Microwave observations provide clues to the origin of Saturn's main rings}}.
\newblock PhD thesis, Cornell University, New York, January 2017.

\bibitem{2019Icar..317..518Z}
Z.~{Zhang}, A.~G. {Hayes}, I.~{de Pater}, D.~E. {Dunn}, M.~A. {Janssen}, P.~D. {Nicholson}, J.~N. {Cuzzi}, B.~J. {Butler}, R.~J. {Sault}, and S.~{Chatterjee}.
\newblock {VLA multi-wavelength microwave observations of Saturn's C and~B rings}.
\newblock {\em Icar}, 317:518--548, January 2019.

\bibitem{2013ApJ...763L..21C}
Bin {Chen}, T.~S. {Bastian}, S.~M. {White}, D.~E. {Gary}, R.~{Perley}, M.~{Rupen}, and B.~{Carlson}.
\newblock {Tracing Electron Beams in the Sun's Corona with Radio Dynamic Imaging Spectroscopy}.
\newblock {\em ApJL}, 763(1):L21, January 2013.

\bibitem{2022ApJ...940..137L}
Yingjie {Luo}, Bin {Chen}, Sijie {Yu}, Marina {Battaglia}, and Rohit {Sharma}.
\newblock {Multiple Regions of Nonthermal Quasiperiodic Pulsations during the Impulsive Phase of a Solar Flare}.
\newblock {\em ApJ}, 940(2):137, December 2022.

\bibitem{2018ApJ...857..133B}
T.~S. {Bastian}, J.~{Villadsen}, A.~{Maps}, G.~{Hallinan}, and A.~J. {Beasley}.
\newblock {Radio Emission from the Exoplanetary System e Eridani}.
\newblock {\em ApJ}, 857(2):133, April 2018.

\bibitem{2018ApJ...866..155P}
J.~Sebastian {Pineda} and Gregg {Hallinan}.
\newblock {A Deep Radio Limit for the TRAPPIST-1 System}.
\newblock {\em ApJ}, 866(2):155, October 2018.

\bibitem{2020ApJ...896...99W}
Brian~E. {Wood}, Samuel {Tun-Beltran}, Jason~E. {Kooi}, Emil~J. {Polisensky}, and Teresa {Nieves-Chinchilla}.
\newblock {Inferences About the Magnetic Field Structure of a CME with Both In Situ and Faraday Rotation Constraints}.
\newblock {\em ApJ}, 896(2):99, June 2020.
     

\bibitem{8879392}
  J.~Flygare, B.~Dong, J.~Yang, L.~Helldner, M.~Dahlgren, J.~Chengjin, M.~Pantaleev, G.~Hovey, and J.~Conway.
  \newblock {Wideband single pixel feed system over 4.6–24 GHz for the Square Kilometre Array}.
  \newblock{\em International Conference on Electromagnetics in Advanced Applications (ICEAA)}, 0630-0635, 2019.

\bibitem{2021AJ....161..111D}
V. {Dike}, M.~R.{Morris},  R.~M.{Rich}, M.~O.{Lewis}, L.~H.{Quiroga-Nu{\~n}ez},  M.~C.{Stroh}, A.~C.{Trapp}, M.~J.{Claussen}.
        \newblock {Ground Vibrational State SiO Emission in the VLA BAaDE Survey}.
      \newblock{\em Astrophysical Journal} journal 161:111, 2021.


\bibitem{2023IJMw....3..570K}
       J.~{Kooi}, M.~{Soriano}, J.~{Bowen}, Z.~{Abdulla}, L.~{Samoska}, A.~{Fung}, R.~{Manthena}, D.~{Hoppe}, H.~{Javadi}, T.~{Crawford}, D.~{Hayton}, I.~{Malo-Gomez}, G.-P.~{Juan Daniel}, A.~{Akgiray}, B.~{Gabritchidze}, K.~A.{Cleary}, C.~S.{Jacobs}, and J.{Lazio}.
        \newblock{A Multioctave 8 GHz‑40 GHz Receiver for Radio Astronomy}.
      \newblock{\em IEEE Journal of Microwaves},
     3:570-586, 2023.

\bibitem{Fconfiguration}
J.~M. {Wrobel} and R.~C. {Walker}.
\newblock {A Suggested Final Configuration for the Very Large Array}.
\newblock {\em arXiv e-prints}, page arXiv:2305.19973, 2023.

\bibitem{ngvla126}
C.~L. {Carilli}.
\newblock {ngVLA Memorandum~126: VLA–ngVLA Transition: Comparison of Two Representative Fixed-Mixed Scale Configurations to the Current VLA}.
\newblock Technical Report 126, National Radio Astronomy Observatory, 2024.

\end{thebibliography}

%\section*{References}
%\printbibliography

\end{document}